\documentclass[a4paper,11pt]{article}
\pdfoutput=1 

\usepackage{jcappub} 

\usepackage[utf8]{inputenc}
\usepackage[T1]{fontenc} 

\usepackage[normalem]{ulem} 
\usepackage{xcolor} 
\usepackage{multirow}
\usepackage{orcidlink}
\title{\boldmath Alcock-Paczynski blinding scheme for the  Ly$\alpha$ forest analysis}

\affiliation{Affiliations are in Appendix \ref{sec:affiliations}}

\author[1]{{G.~Perez-Sanchez}\orcidlink{0009-0009-6473-5368},}
\author[1]{{S.~F.~Beltran}\orcidlink{0000-0001-6324-4019},}
\author[1,2]{{G.~Niz}\orcidlink{0000-0002-1544-8946},}
\author[3]{{S.~Brieden}\orcidlink{0000-0003-3896-9215},}
\author[4,5]{{L.~Verde}\orcidlink{0000-0003-2601-8770},}
\author[4,6]{{A.~Font-Ribera}\orcidlink{0000-0002-3033-7312},}
\author[7]{{J.~Aguilar},}
\author[8]{{S.~Ahlen}\orcidlink{0000-0001-6098-7247},}
\author[9,10]{{D.~Bianchi}\orcidlink{0000-0001-9712-0006},}
\author[11]{{D.~Brooks},}
\author[7]{{T.~Claybaugh},}
\author[7]{{A.~Cuceu}\orcidlink{0000-0002-2169-0595},}
\author[12]{{A.~de la Macorra}\orcidlink{0000-0002-1769-1640},}
\author[13,14]{{B.~Dey}\orcidlink{0000-0002-5665-7912},}
\author[11]{{P.~Doel},}
\author[7,15]{{S.~Ferraro}\orcidlink{0000-0003-4992-7854},}
\author[16,17]{{J.~E.~Forero-Romero}\orcidlink{0000-0002-2890-3725},}
\author[18,19,20]{{E.~Gaztañaga}\orcidlink{0000-0001-9632-0815},}
\author[21]{{S.~{Gontcho A Gontcho}}\orcidlink{0000-0003-3142-233X},}
\author[1]{{A.~X.~Gonzalez-Morales}\orcidlink{0000-0003-4089-6924},}
\author[22]{{G.~Gutierrez},}
\author[23,24]{{H.~K.~Herrera-Alcantar}\orcidlink{0000-0002-9136-9609},}
\author[25,26,27]{{K.~Honscheid}\orcidlink{0000-0002-6550-2023},}
\author[28,29]{{D.~Huterer}\orcidlink{0000-0001-6558-0112},}
\author[30]{{M.~Ishak}\orcidlink{0000-0002-6024-466X},}
\author[31]{{R.~Joyce}\orcidlink{0000-0003-0201-5241},}
\author[7]{{A.~Kremin}\orcidlink{0000-0001-6356-7424},}
\author[11]{{O.~Lahav}\orcidlink{0000-0002-1134-9035},}
\author[7]{{A.~Lambert},}
\author[7]{{M.~Landriau}\orcidlink{0000-0003-1838-8528},}
\author[32]{{L.~Le~Guillou}\orcidlink{0000-0001-7178-8868},}
\author[33,6]{{M.~Manera}\orcidlink{0000-0003-4962-8934},}
\author[25,34,27]{{P.~Martini}\orcidlink{0000-0002-4279-4182},}
\author[4,6]{{R.~Miquel},}
\author[19]{{S.~Nadathur}\orcidlink{0000-0001-9070-3102},}
\author[24,7]{{N.~Palanque-Delabrouille}\orcidlink{0000-0003-3188-784X},}
\author[35,36,37]{{W.~J.~Percival}\orcidlink{0000-0002-0644-5727},}
\author[38]{{F.~Prada}\orcidlink{0000-0001-7145-8674},}
\author[39]{{I.~P\'erez-R\`afols}\orcidlink{0000-0001-6979-0125},}
\author[40]{{G.~Rossi},}
\author[41]{{E.~Sanchez}\orcidlink{0000-0002-9646-8198},}
\author[42]{{E.~F.~Schlafly}\orcidlink{0000-0002-3569-7421},}
\author[7]{{D.~Schlegel},}
\author[28,29]{{M.~Schubnell},}
\author[7]{{J.~Silber}\orcidlink{0000-0002-3461-0320},}
\author[31]{{D.~Sprayberry},}
\author[29]{{G.~Tarl\'{e}}\orcidlink{0000-0003-1704-0781},}
\author[31]{{B.~A.~Weaver},}

\emailAdd{gustavo.perez@ugto.mx}
\emailAdd{gniz@ugto.mx}

\abstract{We present and validate a blinding method for the Lyman-$\alpha$ (Ly$\alpha$) forest analysis based on a modification of the Alcock-Paczynski test. In order to hide the background expansion history, the method employs a geometrical shift of each quasar (QSO) forest in wavelength space, once the quasar continuum has been fitted and the fluctuation field is extracted. The redshift positions for the QSO sample are also changed in a consistent manner. We show that the method remains effective when applied to real data, where contamination from metals and Lyman-$\beta$ is intrinsically mixed with the Lyman-$\alpha$ forest. This limitation is primarily visible in the 1D correlation function, where other blinding strategies can mitigate the effect. To assess its effectiveness, the prescription is tested against a series of datasets of increasing complexity: from idealized low-noise mocks, to realistic DESI year one synthetic datasets, and finally to data from DESI first data release (DR1), using both the auto (Ly$\alpha\times$Ly$\alpha$) and cross (Ly$\alpha\times$ QSO) correlations. We find that the method robustly shifts the BAO peak position from the 3D correlation functions to the expected value for cosmology changes of around 5\% in the matter content, without altering the shape of the posteriors in the model parameters. 
In conclusion, this catalog-level blinding strategy is a viable method for cosmological inference with the Lyman-$\alpha$ forest, particularly if a cross-analysis with other tracers using the same blinding strategy is pursued.}

\begin{document}
\maketitle
\flushbottom

\section{Introduction}
\label{sec:intro}
As cosmological experiments enter an era of increasingly high precision, the robustness and reproducibility of their results have become a bigger concern. In this context, it is important to carefully consider sources of bias that can lead to systematic effects comparable to statistical uncertainty. One such source is confirmation bias, the tendency that we as humans have to favor results that align with prior expectations or established measurements. In order to minimize this risk, blinding procedures, which intentionally modify or hide certain information during data analysis, have been widely used in many areas of research.

There are different approaches to blinding strategies, depending on the nature of the data, the stage of the pipeline at which it is implemented, and the degree of robustness against accidental unblinding. In cosmology, since the early blinding implementation of \cite{Conley}, similar methods were popular in supernovae analyses (see for example \cite{Abbot}) and more recently with the weak lensing probes (see for example \cite{Abbot2,kids} and references therein). For galaxy surveys, some blinding strategies have been widely used, such as in \cite{muir.blinding,brieden,validDR1, validDR2}. In this context, large spectroscopic surveys such as the Dark Energy Spectroscopic Instrument (DESI) provide an ideal setting to implement and test such strategies.

DESI is a stage IV spectroscopic instrument that aims to survey the position of more than 50 million galaxies and quasars, covering approximately one third of the sky \cite{DESIoverview,desi2016a,DESI2016b.Instr, DR1}, to obtain the most precise measurements of the expansion history of the Universe to date, with the goal of shedding light on the nature of dark energy \cite{levi}. The instrument's focus and wide-field correctors lead astronomical light onto a focal plane equipped with 5000 optical fibers. These fibers can simultaneously measure spectra of distant objects \cite{Corrector.Miller.2024, FiberSystem.Poppett.2024, silber}, after careful planning and with an efficient target selection algorithm \cite{SurveyOps.Schlafly.2023, 12,13}, which is based on an optimal tiling of the sky and uses the information on previously selected galaxies and quasars from the DESI Legacy Survey \cite{15,16,17}. Data reduction from the spectrographs \cite{18}, together with a redshift estimating pipeline \cite{19}, have been successfully tested during a survey validation period \cite{20}, which included visual inspection campaigns \cite{21,22} for specific target samples: the Milky Way Survey (MWS, \cite{23,24}), the Bright Galaxy Survey (BGS, \cite{25,26}), Luminous Red Galaxies (LRGs, \cite{27,28}), Emission Line Galaxies (ELGs, \cite{29,30}) and Quasars (QSOs, \cite{31, 32}). Also relevant for this work is DESI first data release \cite{DR1}, which provides the first large-scale dataset from the survey, together with the initial cosmological results derived from these data \cite{Y1-BAO}.

In DESI, an important probe of the cosmological expansion comes from the analysis of high redshift quasars and the absorption of their light by neutral hydrogen in the intergalactic medium (IGM) along the line of sight of those quasars. Since these clouds are at lower redshifts than the quasar, their positions are imprinted as absorption features in the QSO spectrum, appearing to the left of the emission lines in the intrinsic, unabsorbed spectrum of the QSO, known as the QSO continuum. In particular, the absorption associated with the QSO Lyman-$\alpha$ and Lyman-$\beta$ emission lines appears as a collection of absorption features on wavelengths smaller than their corresponding emission peak, forming what is known as the Lyman-$\alpha$ and Lyman-$\beta$ "forests". Their ranges are defined in the QSO rest frame and correspond to 1026 - 1216 \AA\ and 920 - 1020\AA\ respectively. Although absorption of the Lyman-$\alpha$ line dominates the signal in these forests, there are additional contributions from other parts of the QSO spectrum, including metal absorption lines and Lyman-$\beta$ contamination. The traditional analysis attempts to separate the unknown continuum from these contaminants using different methods. This results in a field of fluctuations that traces the matter density field. Their correlations map the expansion history of the Universe for $z > 2$ (see for example some of the latest results in the field with the eBOSS sample \cite{34,35,36} and DESI sample \cite{14,Y1-BAO, DR1, 2024-III, 2024-VI}).

In the present study, we explore a geometrical prescription at the catalog level based on a modified background expansion history. We refer to this approach as an Alcock-Paczynski (AP) blinding scheme. It will be the focus of this study when applied to the analysis of high redshift quasars (QSOs) and their forest, especially within the context of the Dark Energy Spectroscopic Instrument (DESI). We describe a blinding prescription at the catalog level for the Lyman-$\alpha$ analysis of the three-dimensional (3D) correlations to constrain the cosmological expansion. Analysis of the Lyman-$\alpha$ correlations is performed using the public code PICCA \cite{34}, developed within DESI collaboration for Lyman-$\alpha$ forest analyses. We implement our blinding scheme within this framework, as a wavelength shift applied consistently to the field of fluctuations in the forest region, resulting in different inferred expansion histories.

With the goal of designing and validating this blinding scheme, in Section \ref{sec:theory}, we present the Alcock-Paczynski blinding framework and describe its implementation for Lyman-$\alpha$ analyses, including cross correlations with Lyman-$\beta$ and QSOs. We also discuss alternative blinding strategies currently used in DESI analyses. Then, in Section \ref{sec:validation}, we validate the method using progressively more realistic datasets, starting from low-noise synthetic catalogs and culminating with DESI DR1 data. We examine the impact of the blinding on the overdensity field, correlation functions, distortion matrices, and inferred BAO parameters, and finish the section with a discussion on the robustness, limitations and performance of the method. Finally, we present our conclusions in Section \ref{sec:conclusions}. We have also included a more technical appendix, which may prove useful to readers who want to understand the details of the blinding implementation in PICCA (Appendix \ref{sec:PICCA}).

\section{The blinding scheme algorithm}
\label{sec:theory}
In this section, we review the main ideas underlying the blinding algorithm, starting with some general definitions and following the approach introduced in \cite{brieden} for galaxies\footnote{See also the work of \cite{blindingbispectrum} for the equivalent blinding algorithm in the galaxy bispectrum.}. We then describe how these ideas are adapted for the present analysis of Lyman-$\alpha$.

\subsubsection*{AP blinding in galaxy surveys.} 

\begin{figure}[t]
\centering 
\includegraphics[width=.8\textwidth]{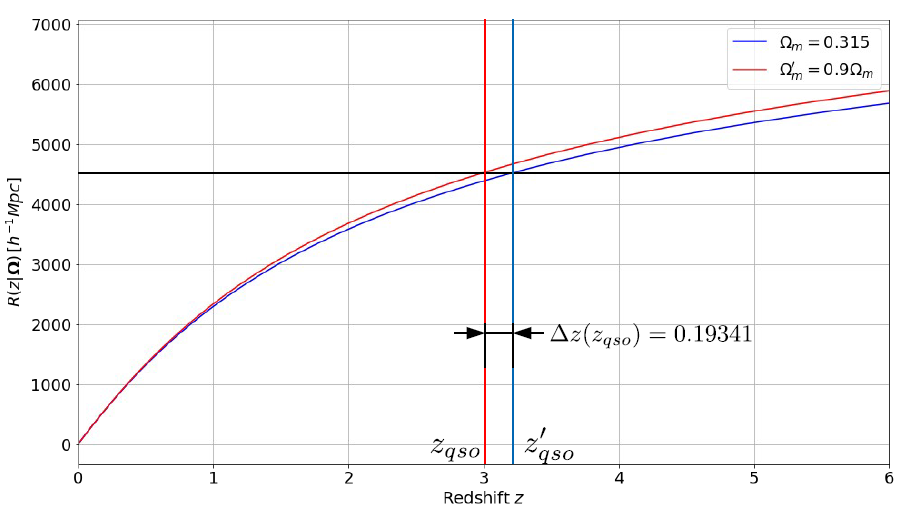}
\caption{Comoving distance for the reference model (blue) and the blinded cosmology (red), with 10\% reduction in the matter density $\Omega_m$ and a fixed $H_0$. The vertical lines show the outcome of matching (horizontal line) the comoving distances for both cosmologies for a single quasar originally at $z\sim 3$, as described in Eq. \eqref{eq:f(z)}.}
\label{fig:zmap}
\end{figure}

In order to hide the true cosmology from observational data, an Alcock-Paczynski (AP) blinding scheme relies on a controlled modification of the expansion history of the Universe. This modification alters the outcome of what is known as the Alcock-Paczynski test (see e.g. \cite{38, 39}), which measures the anisotropic clustering of the sample along the parallel and perpendicular directions to the line of sight (LoS). One way to achieve this is to modify the comoving distance, $R(z)$, between objects in a catalog. This is done by mimicking the evolution under an alternative cosmological model. If a particular cosmological model within a general theory-space is characterized by a set of parameters $\Omega$, the AP blinding scheme directly changes the redshifts of a catalog by matching the comoving distance $R(z|\Omega)$:
\begin{equation}
 z_i\rightarrow R(z'_i|\Omega^{\mathrm{ref}})
 	= R(z_i|\Omega^{\mathrm{bld}})\rightarrow z'_i
\label{eq:Rmaping}
\end{equation}
Here, $\Omega$ refers to the set of parameters that define a given cosmology. The reference (fiducial) cosmology is specified by $\Omega^{\mathrm{ref}}$, while the blinding cosmology is defined by $\Omega^{\mathrm{bld}}$. Figure \ref{fig:zmap} illustrates this matching for two cosmologies. Notice that the alternative choice $R(z_i|\Omega^{\mathrm{ref}}) = R(z'_i|\Omega^{\mathrm{bld}})$ (i.e. exchanging the roles of $z_i\leftrightarrow z'_i$) is not viable, since it would imply a different expansion history at each redshift, as can be seen from the figure.

In the following, we denote shifted variables by primes when referring to the blinded cosmology. At this level, the AP blinding scheme is general enough to map between unblinded cosmologies within a given theoretical framework (e.g. modified gravity). However, given the importance of probing the $\Lambda$CDM paradigm and for simplicity, we focus on a minimal, flat $\Lambda$CDM model, $H^2 = H_0^2\left[\Omega_\Lambda+\Omega_m(1+z)^3\right]$, where only cold dark matter and a cosmological constant are considered (flatness imposes $\Omega_\Lambda=1-\Omega_m$). This is consistent with many Lyman-$\alpha$ analyses in the literature (see e.g. \cite{40} and \cite{41} for the BOSS collaboration). Therefore, one may choose two of the three variables ($H_0$, $\Omega_\Lambda$, $\Omega_m$) to characterize the AP blinding scheme. We choose to vary either $H_0$ or $\Omega_m$, and focus mainly on changes of the form
\begin{equation}
 \Omega'_m = \Omega_m(1+\epsilon)
 \label{eq:gamma_shift}
\end{equation}
where $\epsilon$ is small and $H_0$ is fixed.

The shifts along and across the LoS mentioned above modify the measured BAO peak position, a characteristic feature in the two-point correlation function arising from baryon acoustic oscillations. The BAO peak provides a robust standard ruler that allows for precise measurements of cosmic distances and the expansion history of the Universe. In Lyman-$\alpha$ analysis, it offers one of the few direct and precise probes of the expansion history at high redshifts. For reference, the BAO distance along the angular and redshift directions are
\begin{equation*}
\begin{split}
\Delta\theta\sim r_d/D_M(\bar{z}) \,,
\qquad
\Delta z\sim r_d/D_H(\bar{z}) \,,
\end{split}
\end{equation*}
respectively, where $r_d$ is the sound horizon at the end of the drag epoch and must be calculated for the given cosmology. The argument $\bar{z}$ is the effective redshift of the sample in a given redshift bin, and would also change with the blinding. However, it can be recalculated using the new redshifts, approximated using Eq. \eqref{eq:Rmaping} or left unchanged if the sample is sufficiently homogeneous, such that the number of objects leaving is approximately equal to those entering a given redshift bin \cite{brieden}. The redshift functions $D_H$ and $D_M$ are, respectively, the Hubble distance and the comoving angular diameter distance, given by
\begin{subequations}
\label{eq:ComovDist}
\begin{align}
 D_H(z) & = \frac{c}{H_0\sqrt{1-\Omega_m+\Omega_m(1+z)^3}} \,.
\label{eq:HubDist}              \\
 D_M(z) & = \int_0^z dz' D_H(z') \,,
\label{eq:AngDist}
\end{align}
\end{subequations}

The parameters that characterize the BAO feature's position, the $\alpha$ parameters, are given by
\begin{subequations}
\label{eq:AlphaDef}
\begin{align}
 \alpha_\parallel & = \frac{D_H(\bar{z})/r_d}{\left[D_H(\bar{z})/r_d\right]_{fid}} \,,
\label{eq:a_p}				\\
 \alpha_\perp & = \frac{D_M(\bar{z})/r_d}{\left[D_M(\bar{z})/r_d\right]_{fid}} \,.
\label{eq:a_t}
\end{align}
\end{subequations}
which can be fitted once the correlation functions are calculated, with respect to a previously selected fiducial model (labeled as \textit{fid}). Before proceeding, notice that an AP ratio $\frac{\alpha_\parallel}{\alpha_\perp} = 1$ implies an isotropic clustering of the sample. Moreover, there is an inverse relation between the $\alpha$ values and the position of the BAO peak, such that using a higher value of $\Omega_m$ is expected to shift this feature towards smaller scales.

In the case of a discrete sample with known redshifts, as was done for galaxies in \cite{brieden}, the shift in redshifts reduces to
\begin{equation}
 z' = z+f(z)
\label{eq:f(z)}
\end{equation}
where the smooth function $f = R^{-1}\left(R(z_i|\Omega^{\mathrm{bld}})|\Omega^{\mathrm{ref}}\right)-z$ results from Eq. \eqref{eq:Rmaping} or, equivalently, the $\Delta z$ in Figure \ref{fig:zmap}. We also note that since the comoving distance $R$ is a smooth, monotonic function of z, it is straightforward to compute its inverse $R^{-1}$.

\subsubsection*{Lyman-$\alpha$ forest and its challenges.} 
\begin{figure}[t]
\centering 
\includegraphics[width=.9\textwidth]{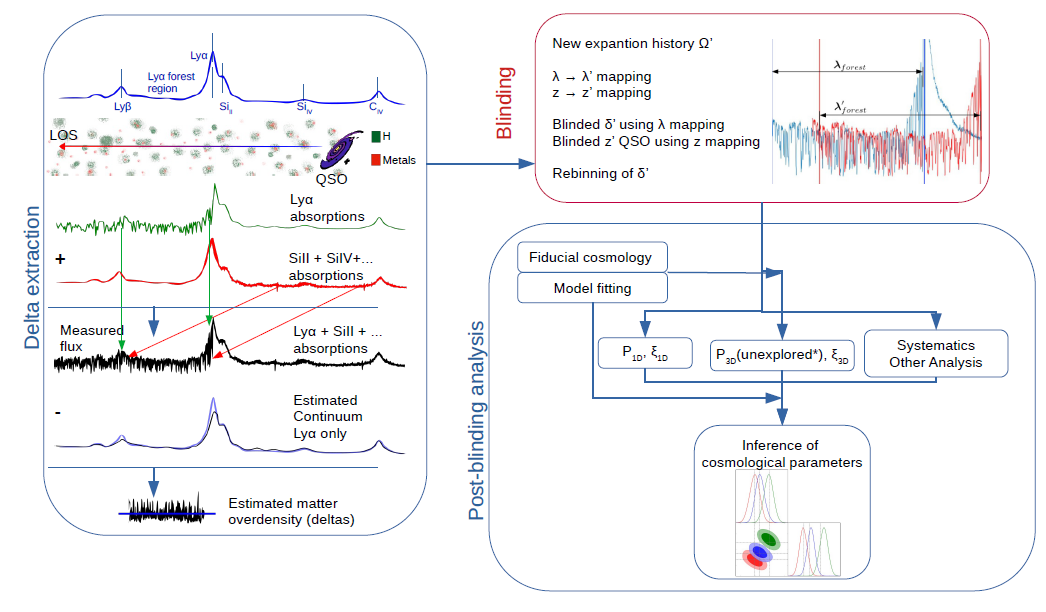}
\caption{Basic steps and key ideas used in the pipeline for Lyman-$\alpha$ forest analysis, including our proposed blinding scheme. The process involves delta extraction from quasar spectra and the application of a blinding transformation to mimic a different expansion history. It then proceeds with post-blinding analysis, including model fitting and systematics checks, to infer cosmological parameters. From the shown possibilities of post-blinding analysis, we focus on the fitting of the correlation functions $\xi_{\mathrm{1D}}$ and $\xi_{\mathrm{3D}}$}
\label{fig:pipeline}
\end{figure}

We now turn to the application of the previous ideas to the Lyman-$\alpha$ forest. In Lyman-$\alpha$ analyses, the relevant observable is the flux fluctuation field, or delta field, which traces the underlying matter density, and is defined as
\begin{equation}
 \delta(\lambda) = \frac{F(\lambda)-\bar{F}(\lambda)}{\bar{F}(\lambda)} = \frac{f(\lambda)}{C(\lambda)\bar{F}(\lambda)}-1
 \label{eq:delta}
\end{equation}
where $F$ is the transmitted flux fraction, $\bar{F}$ its mean value, $f$ the observed flux and $C$ the estimated QSO continuum. A schematic overview of these quantities and the delta extraction process is shown in Figure \ref{fig:pipeline}. As shown in this figure, the observed spectrum from a QSO includes contributions from different physical processes occurring at different redshifts, which makes a direct application of the above procedure insufficient and motivates the adaptations discussed below.

For the AP blinding process to work effectively, we require redshift changes to be applied in a self-consistent way. For spectroscopic catalogs, this translates into controlled wavelength shifts of the observed-frame spectra. In the case of Lyman-$\alpha$ analysis, this introduces an additional complication: the QSO spectrum includes contributions from both the unabsorbed QSO continuum and from absorbing clouds in the intergalactic medium (see figure \ref{fig:pipeline}). Since these components have intrinsically different redshifts, they should be shifted differently under the blinding transformation. However, the delta field extraction process does not assume a cosmology, making a posterior blinding at this stage not meaningful. A direct consequence of these entangled signals is that the earliest implementation of an AP blinding prescription should be done after the fluctuation field (delta) has been computed (see Appendix \ref{sec:PICCA} for details on Lyman-$\alpha$ pipeline done for DESI).

An additional and more important consequence comes from the shifting of the delta field itself, since there is no simple procedure to distinguish contributions from the absorption of different emission lines. As illustrated in Figure \ref{fig:pipeline}, a given wavelength could receive contributions from different absorption processes. For instance, one contribution could arise from Lyman-$\alpha$ absorption by a cloud at a given redshift, while another may come from SiII absorption at a lower redshift. Since the two matter clouds are at distinct redshifts, the AP blinding should shift their contributions differently. In practice, however, there is no simple way to separate these signals. One therefore assumes that all absorption features come from the same emission line in order to perform the blinding. Given its dominance, the natural choice is to assume that all absorption arises from the Lyman-$\alpha$ emission line. As a consequence, this blinding method would naively break down when considering features produced by metal absorptions in the correlation functions. Since we adopt this approach, we must estimate its impact and assess whether the signal-to-noise of a given sample could allow for accidental unblinding based on shifts of the metal peaks in the correlation functions. This is precisely the type of effect the blinding procedure is designed to prevent. It turns out, as we discuss in more detail in Section \ref{sec:robustness}, only in the 1D correlation function are these metal features clearly visible as shifted structures, and could be used to unblind the signal.

\subsubsection*{Adapting AP for Lyman-$\alpha$ forest.} 

To address these issues, we now describe how the AP scheme is modified and implemented for Lyman-$\alpha$ forests. As mentioned above, the AP blinding in the present implementation can be viewed as two separate processes: a shift in the QSO redshifts and a blinding of the forest delta field. The former is identical to the galaxy case \cite{brieden} and relies on Eq. \eqref{eq:f(z)}. For the forests, the blinding process is more involved and requires additional considerations.

The natural space to perform the blinding is wavelength-space. However, mapping redshift to wavelength requires associating each absorption feature with an emission wavelength
\begin{equation}
 \lambda = \lambda_{\mathrm{Ly}\alpha}(1 + z)
 \label{eq:lam_z}
\end{equation}
where $z$ is the redshift of the absorbing cloud. Each absorption feature may contain contributions from different regions of the QSO spectrum, which cannot be separated. Nevertheless, one can proceed by assuming that the only emission wavelength relevant for both the Lyman-$\alpha$ and Lyman-$\beta$ forests is the QSO Lyman-$\alpha$ emission peak at $\lambda_{\mathrm{Ly}\alpha}\sim 1,216$ \AA. Under this assumption, the QSO continuum can be removed using standard techniques (see Appendix \ref{sec:PICCA}), and a straightforward derivation leads to the wavelength-space blinding shift
\begin{equation}
 \lambda' = \lambda+g(\lambda)
\label{eq:g(z)}
\end{equation}
with
\begin{equation}
 g(\lambda) =
 f\left(\frac{\lambda}{\lambda_{\mathrm{Ly}\alpha}}-1\right)
 \frac{\lambda}{\lambda_{\mathrm{Ly}\alpha}}.
\end{equation}
This constitutes a key extension of the original AP blinding framework: while the galaxy-based scheme acts on discrete object redshifts, this approach modifies the wavelength grid to consistently blind the continuous absorption field. Importantly, the wavelength mapping is derived from the same redshift transformation $f(z)$ used for the discrete QSO coordinates, connected through Eq. \eqref{eq:lam_z}. Thus, the QSO redshift and Ly-$\alpha$ forest wavelength transformations are not independent operations, but represent the same radial coordinate transformation in redshift and wavelength coordinates, respectively. This ensures that QSOs and Ly-$\alpha$ absorption pixels are assigned consistent blinded coordinates when constructing Ly-$\alpha\times$QSO cross-correlations.

In practice, the Lyman-$\alpha$ analysis involves multiple absorption regions and the AP blinding must therefore be applied to all relevant catalogs: the Lyman-$\alpha$ and Lyman-$\beta$ absorption forests and the QSOs. For the forests, the comoving distance at each point is modified according to the blinded cosmology. For the QSOs, the redshift is shifted accordingly. The validity of the blinding can then be tested in different ways. If all catalogs are modified coherently, the Ly$\alpha\times$QSO cross-correlation should exhibit a BAO peak shift consistent with that of the auto-correlations. In addition, the correlation functions can be decomposed into a smooth component and a BAO peak contribution, $\xi = \xi_{\mathrm{sm}}+\xi_{\mathrm{peak}}$, as shown in \cite{34}. A similar decomposition of the forest$\times$QSO cross correlation can be done with a dip instead of a peak. Since the AP blinding modifies the inferred distances, it is expected to shift the BAO peak while leaving the smooth component largely unaffected. This provides an additional consistency check for the implementation.

\subsection{Step by step blinding algorithm}
\label{sec:algorithm}
\begin{figure}[t]
\centering 
\includegraphics[width=.8\textwidth]{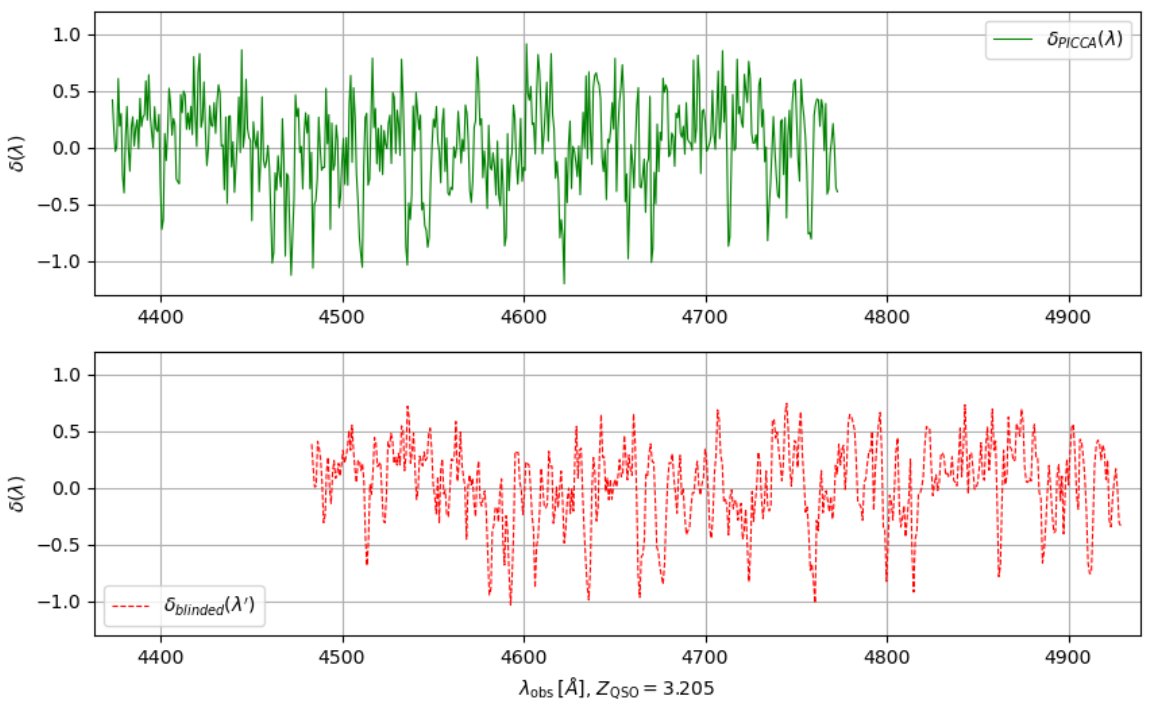}
\caption{Blinding of the $\delta(\lambda)$ field for a single QSO at z = 3.2, using $\epsilon = -0.05$ ($\Omega'_m < \Omega_m$), which results in larger comoving distances. As a result, a larger BAO scale is expected. From visual inspection, one can see that the peaks on the blinded case (red dotted line) lost their "sharpness". This is a result of the rebinning from the blinded wavelength grid to a linearly spaced grid, step necessary for future parts in the pipeline and to avoid unblinding.}
\label{fig:blinddelta}
\end{figure}

In summary, the AP blinding process can be viewed as the following algorithm:
\begin{enumerate}
 \item \textbf{Pre-calculated blinding map.} Construct maps, using the functions defined in Eqs. \eqref{eq:f(z)} and \eqref{eq:g(z)}, to coherently shift redshifts and wavelengths that relate two different expansion histories, defined by parameters $\Omega^{\mathrm{bld}}$ (blinded cosmology) and $\Omega^{\mathrm{ref}}$ (reference cosmology). The reference cosmology may or may not coincide with the true background evolution of the data, but is expected to be close to it. The redshift blinding map $f(z)$ is constructed by matching the comoving distance $R(z|\Omega)$ between the two cosmologies, as in Eq. \eqref{eq:f(z)}. Substituting $f$ in the expression for $g$ (Eq. \eqref{eq:g(z)}) leads to the blinding function for the wavelengths. Since both maps are smooth functions, one may use a few points to describe them and take an interpolation for further accuracy on each application. Moreover, these maps are constructed once, with only their domain and range adjusted to the catalog being blinded.
 
 \item \textbf{Iterate} the following steps for each QSO after extraction of the delta field:
 \begin{enumerate}
  \item \textbf{Redshift blinding.} Shift the QSO redshifts using the previously calculated function $f$, interpolated around the value $z_{\mathrm{qso}}$ for greater precision. This shift is required not only for forest$\times$QSO cross-correlation, but also to correctly transform the observed data to its rest frame, if needed.
  \item \textbf{Delta blinding.} Using the function $f$, calculate the wavelength shift $g$ and apply it to each delta in the forest(s) of the QSO. Figure \ref{fig:blinddelta} illustrates this process for a representative delta field in our catalog.
  \item \textbf{Forest range preservation (optional).} Because the wavelength mapping is non-linear, the full forest range shrinks (expands) for $\epsilon > 0$ ($\epsilon < 0$). If required, one may correct for this effect. For larger $\Omega'_m$, the empty region at the start of the forest (in wavelength-space) can be filled with Gaussian noise, while for smaller $\Omega'_m$ the excess data can be removed. Notice that a change in $\Omega_m$ of $\sim5\ \%$ causes a comparable change in the forest range for QSOs around $z \sim 2.5$. In the results presented here, however, we choose not to apply this correction.
 \end{enumerate}
 
 \item \textbf{Blinded catalog.} After application of the non-linear mapping, the blinded delta field no longer lies on a homogeneous grid. To prevent unblinding through comparisons of the bin spacing, the delta field and its corresponding weights, which encode the statistical reliability of each forest pixel, are resampled onto a linear array in the chosen scale (e.g., a linear scale in comoving distance for DESI or logarithmic in wavelength for BOSS). The resulting blinded delta files, their resampled weights, and the consistently shifted QSO redshifts are then stored for subsequent analyses.
 
\end{enumerate}

In this work, the blinding is applied at the level of the delta field, without subsequent re-optimization of the weights. While weight optimization can be part of the extraction procedure, we adopt this conservative choice to avoid introducing additional changes beyond the transformation itself.

\subsection{Alternative blinding schemes}
\begin{figure}[tbp]
\centering
\includegraphics[width=.8\textwidth]{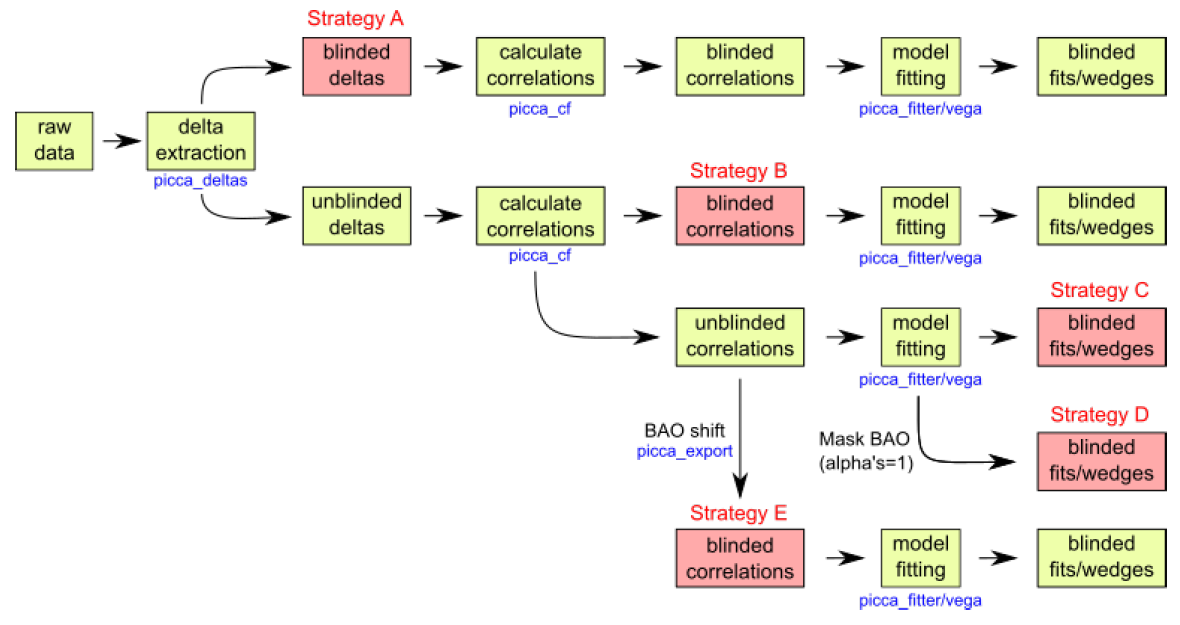}
\caption{Different blinding strategies and their location within the DESI Lyman-$\alpha$ analysis pipeline. Strategy A corresponds to the catalog-level AP blinding scheme explored in this work, while Strategies B--E apply the blinding at progressively later stages of the analysis, either at the level of the measured correlations or during parameter inference.}
\label{fig:Strategies}
\end{figure}

Alternative approaches to blinding can be implemented at later stages of the analysis pipeline, either at the level of the measured correlation functions or directly at the parameter-inference stage. Figure \ref{fig:Strategies} summarizes several possible implementations within the DESI Lyman-$\alpha$ analysis workflow. Strategy A corresponds to the catalog-level AP blinding scheme developed in this work, where the transformation is applied directly to the forests before the computation of correlation functions. In contrast, Strategies B--E apply the blinding after the delta extraction step, either by modifying the measured correlations, altering the fiducial model used in the fit, or shifting the inferred BAO parameters at the posterior level. Among these alternatives, Strategy E is currently the commonly adopted approach in DESI analyses \cite{Y1-BAO,DESI:2025zpo,Cuceu:2025nv}. Implementations of several of these approaches are available in the public PICCA code\footnote{\href{https://github.com/igmhub/picca}{https://github.com/igmhub/picca}} \cite{34}.

Although these later-stage approaches successfully hide the final cosmological constraints, they provide weaker protection against confirmation bias during pipeline development or while using other codes that are not PICCA for calculating the correlation function or model fitting. In addition, some of these approaches may become difficult to generalize when combining multiple datasets or tracers, particularly if AP-like transformations are applied independently to different observables and consistency between the mappings must be maintained.

In contrast, the strategy explored in this work applies the blinding at an earlier stage of the analysis chain, immediately after the extraction of the delta field. This allows the full analysis pipeline to be developed and validated using blinded data, including correlation-function estimation, covariance validation, systematic treatments, and parameter fitting. As a result, the method provides stronger protection against experimenter bias while preserving compatibility with the standard cosmological inference framework.

\section{Validation and performance of the AP blinding scheme}
\label{sec:validation}

An important aspect of the blinding of the Lyman-$\alpha$ forest analysis is to identify where cosmological assumptions enter the analysis pipeline. The delta field extraction relies only on continuum estimation, which can be done using different methods (e.g. \cite{37}) that do not assume cosmological knowledge. After extracting the delta field, correlation functions can be calculated, either 3D correlations in configuration space or 1D correlations in both configuration and Fourier spaces. In this work, we focus on 3D correlations, which have traditionally been used to constrain the BAO scale from the Lyman-$\alpha$ forest. At this stage, a cosmological model is required to convert redshifts and angular separations into comoving distances. In our framework, these distances are computed using a fixed baseline cosmology, which is unchanged by the blinding procedure. We apply the same analysis pipeline to both blinded and unblinded catalogs, using this baseline cosmology for the distance conversion in all cases.

When fitting and comparing a model to those 3D correlations, parameters are estimated relative to a fiducial cosmology. We apply the AP blinding after the delta extraction and leave the rest of the pipeline unchanged (including the distance conversion and the fitting procedure). As a result, the inferred distances become inconsistent with the true (blinded) cosmology, producing the desired AP-like distortion of the correlation function. When fitting data, this distortion is captured by the BAO scaling parameters defined with respect to a fiducial model, therefore hiding the true cosmology of the data.

\subsection{Validation datasets and analysis setup}
\label{sec:catalogs}

Our validation follows a progressive approach, moving from idealized mocks to increasingly realistic catalogs, and finally to the first DESI Data Release (DR1).

For the first catalog, we use simulated datasets generated by a set of codes within the DESI collaboration. The process starts with an initial Gaussian random field, which is transformed into skewers using the CoLoRe suite \cite{42}, avoiding the use of computationally expensive N-body or hydrodynamical simulations. The LyaCoLoRe code \cite{43} then converts CoLoRe’s output into skewers of transmitted flux fraction for Lyman-$\alpha$, with the option to include additional absorption contributions such as Lyman-$\beta$ and metal lines. Finally, the quickquasars package within desisim\footnote{\href{https://github.com/desihub/desisim}{https://github.com/desihub/desisim}} \cite{desisim}, is used to add QSO continua and instrumental noise to each skewer spectrum.

To validate our blinding model, we first use mocks with over one million QSOs, distributed with an angular density of 118 per square degree, with redshift $z>2.1$, and a very high exposure time (1,000 exposures per QSO) to make the noise contribution negligible. We refer to this QSO sample as catalog A when only Lyman-$\alpha$ absorption is included, and catalog B when the Lyman-$\beta$, SiII(1190), SiII(1193), SiIII(1207) and SiIII(1260) absorption lines are incorporated.

As an intermediate step, we also consider a more realistic mock catalog, denoted catalog C, including realistic observational noise together with Lyman-$\alpha$, Lyman-$\beta$ and metal absorptions. This catalog contains approximately 420,000 QSOs with one exposure per object and was designed to broadly reproduce early expectations for DESI DR1 prior to the availability of finalized DR1 mock suites. As such, this catalog serves as an intermediate step between the idealized low-noise mocks and the final DESI DR1 analysis.

Lastly, we apply the blinding to DESI DR1 data. A more extensive validation using larger and more recent mocks is left for future work. Nevertheless, the range of noise levels and data complexities considered here already provides a broad validation regime for assessing the robustness of the method.

The AP parameters, defined in Eqs. \eqref{eq:AlphaDef}, describe the BAO peak position along and perpendicular to the LoS. Since the reference cosmology is chosen to match that of the mock catalogs, in an ideal fitting process the AP parameters are expected to be equal to unity in the absence of blinding.

In practice, the measurement of these parameters is affected by finite-sample fluctuations, leading to deviations from unity even in the absence of blinding. Since the blinded and reference measurements are derived from the same underlying catalog, we compare both results through the normalized parameters
\begin{equation}
 q_i = \frac{\alpha_i^{\mathrm{bld}}}{\alpha_i^{\mathrm{ref}}},
 \label{eq:qdef}
\end{equation}
where the subindex $i = \parallel, \perp$ runs over the clustering with respect to the LoS, while the superscript denotes if the data is unaltered (ref) or blinded (bld).

We quantitatively compare the measured values of $\alpha_\parallel$ and $\alpha_\perp$ with the expected values computed from the Hubble distance and the comoving angular diameter using CLASS \cite{44}. For the fitting, we adopt models and parameter choices from previous research on synthetic datasets \cite{45} and eBOSS data \cite{34}. These choices reflect the fact that the mock catalogs used in this work were constructed within an eBOSS-like framework, ensuring consistency between the fitting model and the simulated data. In particular, we use the power spectrum model of \cite{46}, including non-linear corrections following \cite{47}. For the application to DESI DR1 data, we adopt the same fitting framework and parameter choices used in the DESI DR1 analysis \cite{Y1-BAO}, allowing a direct comparison of the blinding performance across datasets.

Posterior sampling is performed using Polychord \cite{48,49}, together with Vega\footnote{\href{https://github.com/andreicuceu/vega}{https://github.com/andreicuceu/vega}}, which implements the modelling and fitting framework. In our analysis, only the BAO peak position parameters are varied, while the remaining nuisance parameters are fixed to those of previous eBOSS analyses \cite{34,35,45,50}. This choice isolates the geometric effect of the blinding on the BAO scale and avoids degeneracies with nuisance parameters, which are not the focus of this study. The fixed parameters remain the same for the reference and blinded cases.

\subsection{Effects of blinding on the overdensity field}
\label{sec:effectsondelta}

An example of the AP blinding on a single QSO forest was shown in Figure
\ref{fig:blinddelta}, where the field is shifted to the right due to the
smaller chosen value for $\Omega_m$. As a result of the chosen
$\epsilon = -0.05$ and the particular QSO redshift, the range of the
forest is enlarged by $\sim 50$ \AA.

\begin{figure}[t]
\centering
\includegraphics[width=.8\textwidth]{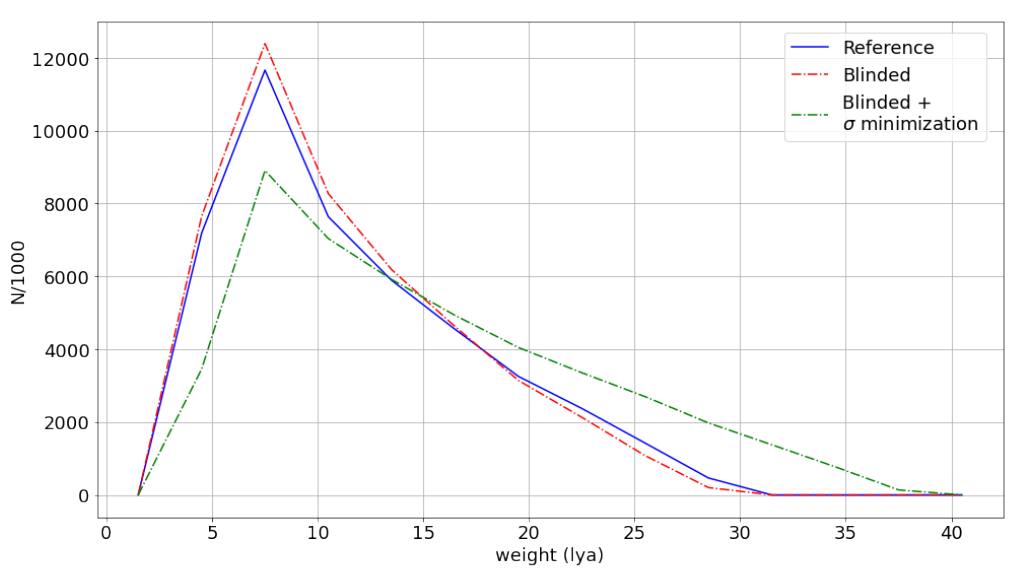}
\caption{Histogram for the delta weights for a simulation of Y1 DESI survey data with only Lyman-$\alpha$ absorptions. The total area over the histogram is conserved by the blinding. There is a tendency to lower weights for the blinding algorithm. This is a consequence of the loss of fine detail in the peaks of the overdensity, due to the resampling of the forest to a linear $\log\lambda$ grid. The green line corresponds to a variation in which the blinding is applied before the last weight iteration of Eq. \eqref{eq:varianceiterations}, which shows unrealistic changes towards higher weights and is therefore discarded.}
\label{fig:weights}
\end{figure}

We also find that the blinded overdensity field has slightly different weights associated with each wavelength (see Figure \ref{fig:weights}). This mild weight modification is also found for other choices of the blinding parameters. Both the loss of sharpness of some peaks and the weight modifications are due to the rebinning process at the end of the algorithm, which is necessary to prevent unblinding from comparisons of the $\lambda$-grid and to maintain a linear spacing in the data. As such, these effects are not easy to prevent. To characterize the impact of the resampling procedure, we tested several interpolation methods and flux-conserving resampling schemes\footnote{We tested several interpolation methods available within \href{https://docs.scipy.org/doc/scipy/tutorial/interpolate.html}{SciPy}, as well as the flux-conserving resampling method implemented in \href{https://desispec.readthedocs.io/en/latest/_modules/desispec/interpolation.html}{desispec}. The final implementation uses linear interpolation via \texttt{numpy.interp}.} and found no significant differences in the resulting correlation functions or inferred BAO parameters. The inverse variance is appropriately transformed to account for the change in grid spacing. We also find similar mild rebinning effects for different choices of the blinding parameters, with no evidence for a significant dependence on the magnitude of the blinding shift over the range explored in this work.

As a sanity check, we also investigated the case in which the blinding is applied before the last weight iteration of Eq. \eqref{eq:varianceiterations}, but found that it leads to an unrealistic change towards higher weights that do not correspond to a physical lower variance in the data, but rather to artifacts of the minimization procedure. As such, we discard this methodological variant. While the resampling conserves the overall flux and weighting of the data, as shown by the area conservation under the curves in Figure \ref{fig:weights}, the impact of these modifications is ultimately assessed at the level of correlation functions and BAO parameters inference, where we observe no significant changes in the recovered parameters or their uncertainties. Hence, we conclude that this is a mild and acceptable artifact.

\subsection{Validation using synthetic catalogs}
\label{sec:lownoisevalidation}

To study the effect of the blinding on the BAO peak, we first consider the $\mathrm{Ly}\alpha(A)$ auto-correlation in noiseless mocks without metals, and then extend the analysis to more realistic scenarios.

\subsubsection*{Catalog A: noiseless Ly$\alpha$ mocks}
\begin{figure}[t]
\centering 
\includegraphics[width=.8\textwidth]{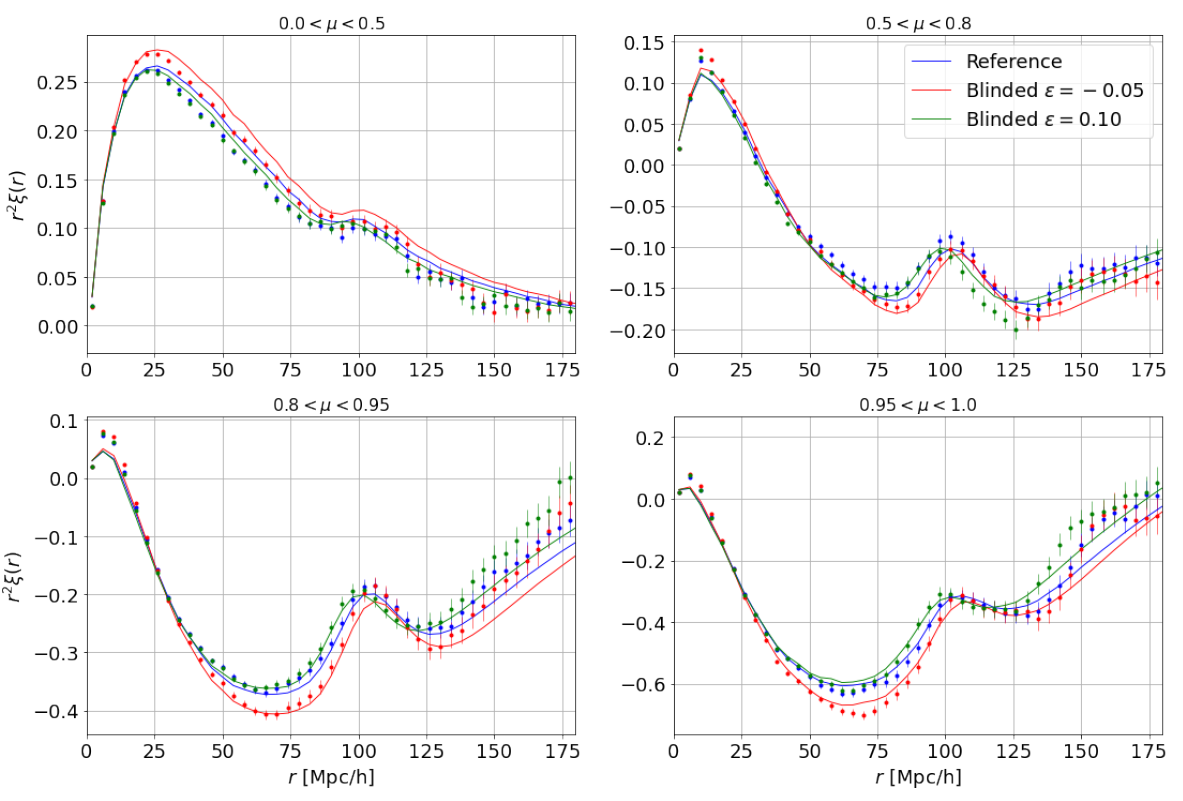}
\caption{$\mathrm{Ly}\alpha(\mathrm{Ly}\alpha)$ auto-correlation function for catalog A (low noise and no metals) for four different wedges. Blue markers represent the unaltered correlation function, while red and green show the blinded cosmology with $\Omega'_m = 0.95\Omega_m$ ($\epsilon = -0.05$) and $\Omega'_m = 1.1\Omega_m$ ($\epsilon = 0.1$), respectively. The solid lines correspond to a $\chi^2$ minimization using the PICCA-fitter, which we only use for visualization of the model.}
\label{fig:mockAcorrs}
\end{figure}

\begin{figure}[t]
\centering 
\includegraphics[width=.8\textwidth]{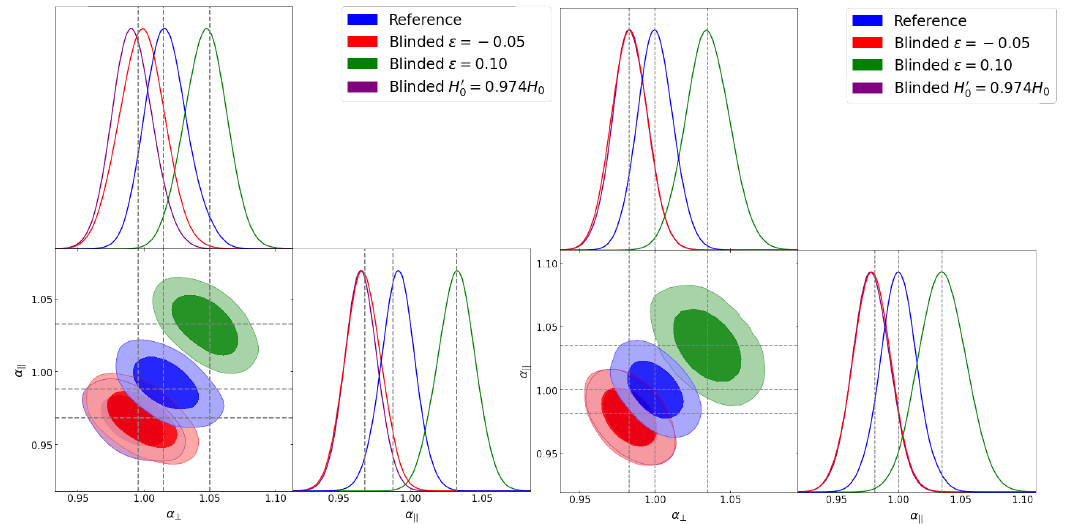}
\caption{Posteriors of the BAO peak parameters $\alpha_\parallel$ and $\alpha_\perp$, using the $\mathrm{Ly}\alpha(A)$ auto-correlation (left) and $\mathrm{Ly}\alpha(A)\times \mathrm{QSO}$ cross-correlation (right), for the low noise catalog. In all cases, the recovered values are within one sigma of the expected values. The $H_0$ modification is selected to match the $\epsilon = -0.05$ expected values, to show that the two approaches lead to consistent shifts in the BAO parameters within the statistical uncertainties. The shift in the $\alpha$-parameters is consistent with theoretical predictions.}
\label{fig:mockAposts}
\end{figure}

We first show the effect of two different values of the blinding parameter: $\epsilon = -0.05$ and $\epsilon = 0.1$, which result in opposite direction shifts of the BAO in the 3D $\mathrm{Ly}\alpha(A)$ auto-correlation. As shown in Figure \ref{fig:mockAcorrs}, the BAO peak is shifted to the right (left) for $\epsilon = -0.05$ ($\epsilon = 0.1$), consistent with a smaller (larger) matter density. These induced shifts are correctly captured by the model fit to the correlation, as appreciated in the posteriors of Figure \ref{fig:mockAposts}, where the recovered BAO inferred parameters are consistent with the expected values from the blinded cosmology.

In addition to the BAO peak shift, we observe a number of small effects associated with the blinding procedure. We note a mild mismatch between the best-fit model and the data at low $\mu$, particularly in the overall shape and amplitude of the correlation function. This is expected, as only the BAO parameters are varied in the fit while nuisance parameters are kept fixed. As a result, the model is not optimized to reproduce the full shape of the correlation function.

We also observe small changes in the posterior widths for larger values of the blinding parameter (e.g. $\varepsilon = 0.1$). This is primarily due to the remapping and rebinning of the forest, which slightly modifies the distribution of pixels and their associated weights.

As discussed in Section \ref{sec:theory}, one may alternatively blind the data through a modification of the Hubble parameter. For example, the choice $H'_0 = 0.974H_0$ produces posterior contours very similar to those obtained for $\epsilon = -0.05$, as evident in Figure \ref{fig:mockAposts}. A small shift is visible, particularly in $\alpha_\perp$, which we attribute to statistical fluctuations and to weak correlations introduced by the blinding transformation. Taken together, these effects remain subdominant and do not affect the recovery of the BAO peak position.

The $\mathrm{Ly}\alpha(A)\times \mathrm{QSO}$ cross-correlation provides an additional test of the internal consistency of the blinding method. Since the Lyman-$\alpha$ forest and the QSO positions are blinded independently, any disagreement would produce a broadening or distortion of the BAO peak. Instead, we find that the BAO parameters match the expected shifted values and that the posterior widths show no significant broadening. The relative changes in the posterior widths are below 0.5\% for both $\alpha_\parallel$ and $\alpha_\perp$.

\subsubsection*{Catalog B: inclusion of Ly$\beta$ and metal absorptions}
\label{sec:low noise}

\begin{figure}[t]
\centering 
\includegraphics[width=.8\textwidth]{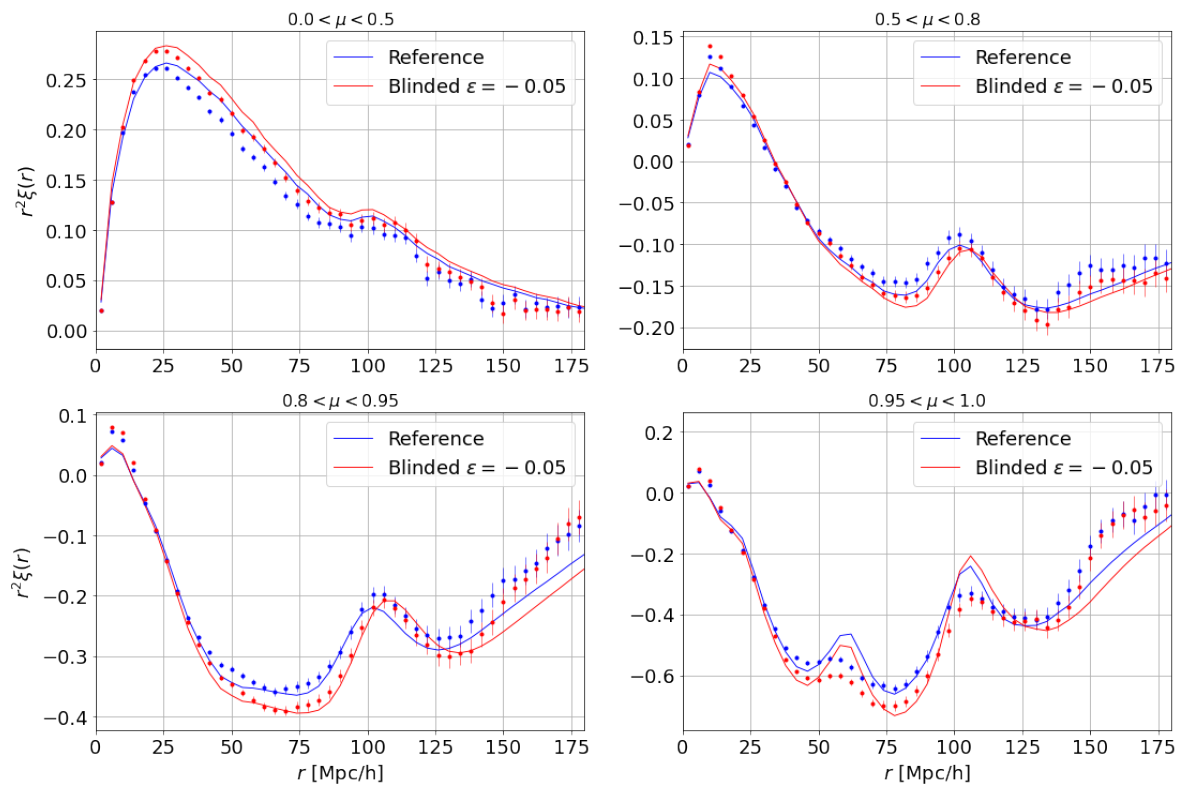}
\caption{$\mathrm{Ly}\alpha(\mathrm{Ly}\alpha)$ auto-correlation function and fitting of the $\alpha_\parallel$, $\alpha_\perp$ parameters for catalogue B (low noise with Lyman-$\alpha$, Lyman-$\beta$ and SiII, SiIII lines). Solid lines correspond to a $\chi^2$ minimization using the PICCA-fitter. Notice the displacement of the BAO peak position related to the metal bump; the BAO peak for the blinded correlation (red) is clearly shifted, while a shift of the metal bump is not visually perceived.}
\label{fig:mockBcorrs}
\end{figure}

We now extend the analysis to catalog B, which includes Lyman-$\beta$ and metal absorption lines.
The wedges of the 3D $\mathrm{Ly}\alpha(A)$ auto-correlation function for the B catalog are shown in Figure \ref{fig:mockBcorrs}. The BAO peak remains clearly shifted according to the imposed cosmology, while the metal-induced features show a weaker apparent displacement. This occurs because the blinding map assumes that all absorption arises from Lyman-$\alpha$, causing metal features, associated with different rest-frame wavelengths, to be assigned incorrect distances. As before, the best-fit model is not expected to fully match the metal-induced features, since only BAO parameters are varied in the fit.

\begin{figure}[t]
\centering 
\includegraphics[width=.8\textwidth]{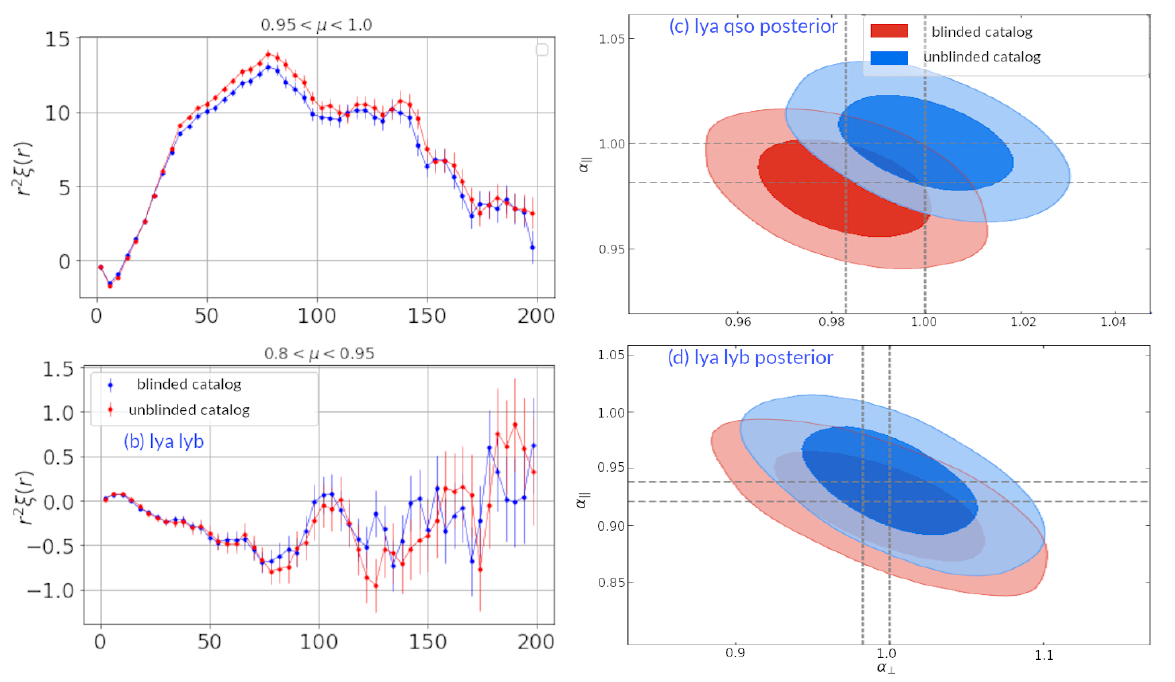}
\caption{Left column: Wedges for the $\mathrm{Ly}\alpha(\mathrm{Ly}\alpha)\times \mathrm{QSO}$ (top) and $\mathrm{Ly}\alpha(\mathrm{Ly}\alpha)\times \mathrm{Ly}\alpha(\mathrm{Ly}\beta)$ (bottom) correlation functions for catalog B (low noise with Lyman-$\alpha$, Lyman-$\beta$ and SiII, SiIII lines). Solid lines show the best PICCA-fitter model. Right column: posteriors of the BAO position parameters for the same correlation cases as in the left column.}
\label{fig:mockBposts}
\end{figure}

Nevertheless, the posterior widths and central values of the BAO parameters remain consistent with the expected shifts, as shown in Figure \ref{fig:mockBposts}\footnote{For this figure, we use the PICCA-fitter as implemented in earlier stages of the analysis; the main results of this work are obtained using VEGA and are not affected by this choice.}. This agreement is observed not only for the $\mathrm{Ly}\alpha(A)$ auto-correlation, but also for the $\mathrm{Ly}\alpha(A)\times \mathrm{Ly}\alpha(B)$ and $\mathrm{Ly}\alpha(A)\times \mathrm{QSO}$ correlations.

\subsubsection*{Combined probes and summary of recovered parameters}
Table \ref{tab:qresults} summarizes the measured and expected values of the normalized BAO parameters for different catalogs and blinding parameters. In all cases, the measured values are consistent with the expected ones within one sigma, including the larger $\epsilon = 0.1$ shift. No significant degradation of the recovered BAO parameters is observed over the range of blinding shifts tested.

\begin{figure}[tbp]
\centering 
\includegraphics[width=.9\textwidth]{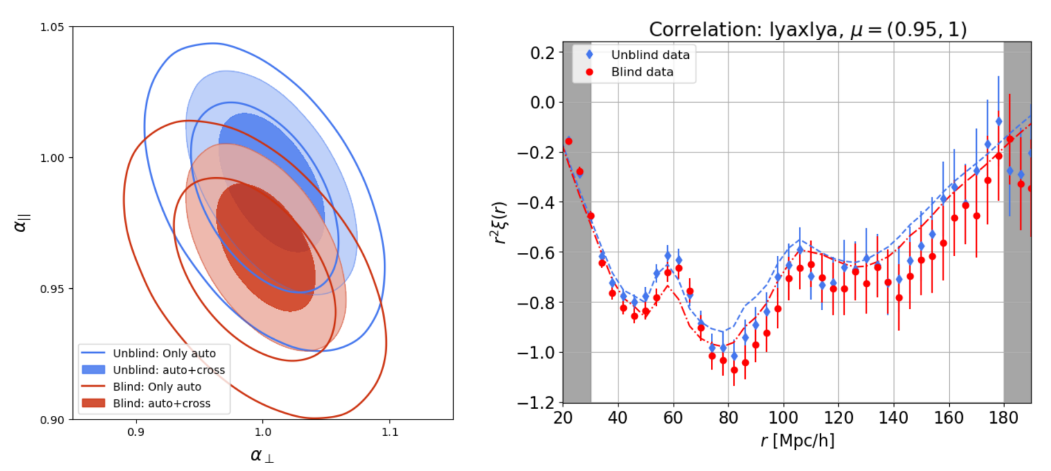}
\caption{Left: Posterior contours for the $\alpha_\parallel$ and $\alpha_\perp$ parameters, using correlations internal to the Lyman-$\alpha$ forest (i.e., combining the $\mathrm{Ly}\alpha$(A) and $\mathrm{Ly}\alpha$(B) regions) together with the $\mathrm{Ly}\alpha(A)\times \mathrm{QSO}$ cross-correlation. All datasets taken from catalog C (see Section \ref{sec:catalogs}). The similar widths of the contours demonstrate that our blinding procedure preserves the ability to combine different datasets, provided a consistent transformation is applied to all of them. A small rotation of the contours is observed in the blinded combined case, consistent with weak correlations introduced by the blinding. Right: Autocorrelation functions (and their best fit) associated to those contours.}
\label{fig:combined}
\end{figure}

We also performed parameter inference on catalog C, combining the $\mathrm{Ly}\alpha(A)$ auto-correlation and the $\mathrm{Ly}\alpha(A)\times \mathrm{QSO}$ cross-correlation, with an example result shown in Figure \ref{fig:combined}. The posterior shapes remain largely unchanged, although a small tilt is observed, consistent with weak correlations introduced by the blinding transformation. This effect is small and does not impact the recovery of the BAO parameters, whose central values remain consistent with expected shifted cosmology.

This agreement between auto- and cross-correlations demonstrates that the blinding method can be consistently applied across different datasets, provided the same AP transformation is used. This opens the possibility of extending the approach to combined analyses including additional DESI tracers under a common blinding scheme.

\begin{table}[tbp]
\centering
\begin{tabular}{|l | l | c | c | c | c | c | c|}
    \hline
    Correlation & Blinding shift & \multicolumn{3}{c|}{$q_{\parallel}$} & \multicolumn{3}{c|}{$q_{\perp}$} \\
    \cline{3-8}
    &  & Meas. & Error & Exp. & Meas. & Error & Exp. \\ \hline
   Cat A: Ly$\alpha \times$ Ly$\alpha$ & $\epsilon = 0.1$ & 1.041 & 0.036 & 1.035 & 1.031 & 0.033 & 1.035 \\ \hline
    Cat A: Ly$\alpha \times$ QSO & $\epsilon = 0.1$ & 1.035 & 0.037 & 1.035 & 1.035 & 0.036 & 1.035 \\ \hline
    Cat A: Ly$\alpha \times$ Ly$\alpha$ & $\epsilon = -0.05$ & 0.974 & 0.034 & 0.980 & 0.983 & 0.033 & 0.983 \\ \hline
	Cat A: Ly$\alpha \times$ QSO & $\epsilon = -0.05$ & 0.978 & 0.032 & 0.980 & 0.983 & 0.032 & 0.983 \\ \hline
	Cat A: Ly$\alpha \times$ QSO & $\epsilon_{H_0} = -0.026$ & 0.978 & 0.031 & 0.980 & 0.983 & 0.031 & 0.983 \\ \hline
	Cat B: Ly$\alpha \times$ Ly$\alpha$ & $\epsilon = -0.05$ & 0.973 & 0.033 & 0.980 & 0.976 & 0.032 & 0.983 \\ \hline
	Cat B: Ly$\alpha \times$ QSO & $\epsilon = -0.05$ & 0.981 & 0.026 & 0.980 & 0.986 & 0.026 & 0.983 \\ \hline
\end{tabular}
\caption{Fitted blinding parameters $q_i$ (see Eq. \eqref{eq:qdef}) for the low-noise catalogs A (no metals) and B (with metals) using the auto- and cross- correlation functions. The blinding shifts are defined either through $\epsilon_{H_0} = H'_0/H_0 - 1$ or through Eq. \eqref{eq:gamma_shift}. Uncertainties are computed assuming uncorrelated variables. The expected (Exp.) values are calculated from the input (blinded) cosmology (analytic or CLASS), while the measured (Meas.) values are obtained from the fits. In all cases, values are consistent within the reported uncertainties.}
\label{tab:qresults}
\end{table}

\subsection{Validation on DESI DR1 data}
\label{sec:testsY1}

We now turn our focus to real data from DESI DR1. This dataset, which we denote catalog Y1, includes spectra from more than 420,000 quasars with redshift $z>2.1$ and provides a benchmark against which we validate the performance of the blinding methodology, through comparison with the published DESI DR1 results \cite{Y1-BAO}.

Before extracting the $\delta(\lambda)$ fields, we account for known contaminants following the same procedure as DESI's main analysis of DR1 data \cite{Y1-BAO}, including application of masks for DLAs, BALs and sky lines, together with standard quality cuts.

\begin{figure}[t]
\centering 
\includegraphics[width=.8\textwidth]{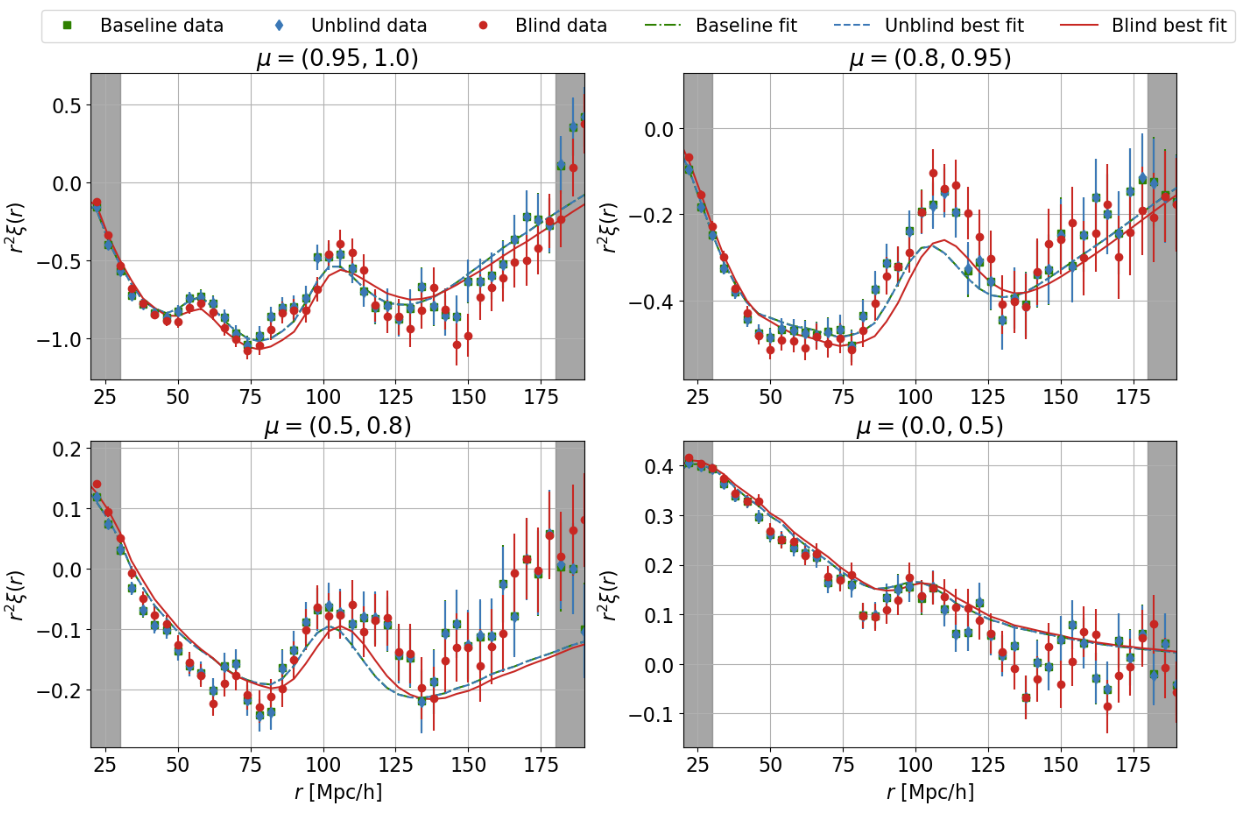}
\caption{Ly$\alpha$(A) auto-correlation function for Y1 catalog. Baseline analysis corresponds to the one reported in \cite{Y1-BAO}, added here to show that there is no noticeable difference between it and our unblinded Y1 catalog. The blinded Y1 catalog shows the expected shift, consistent with what was found for catalogs A and B.}
\label{fig:Y1corr}
\end{figure}

\begin{figure}[t]
\centering 
\includegraphics[width=.6\textwidth]{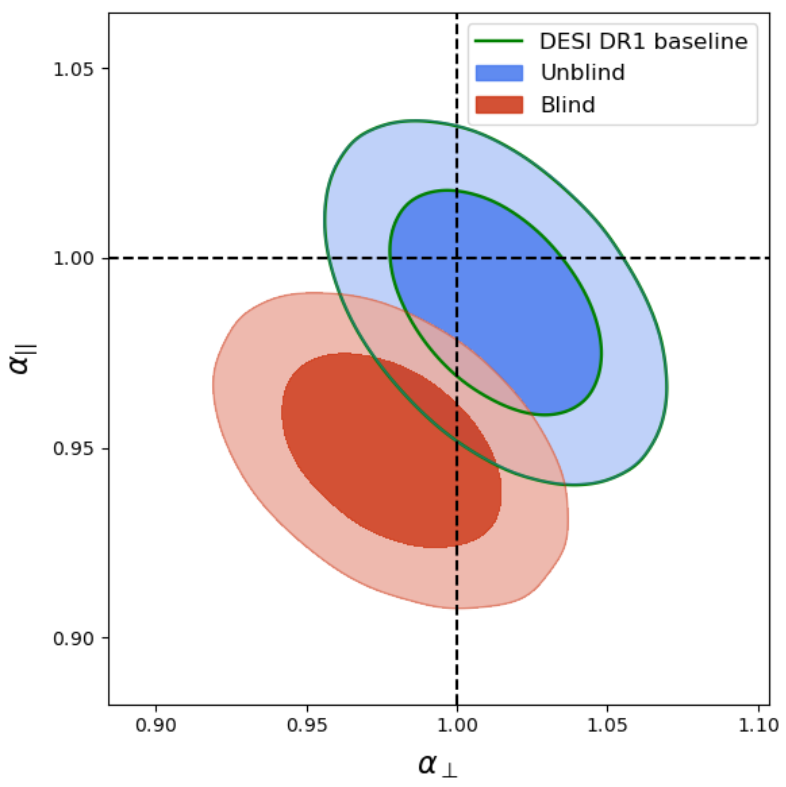}
\caption{Posteriors of the BAO parameters, computed with VEGA and including both auto- and cross-correlation functions. Our results are consistent with those obtained for the synthetic catalogs. Furthermore, the unblinded results are fully consistent with the DESI DR1 Lyman-$\alpha$ BAO baseline reported in \cite{Y1-BAO}.}
\label{fig:Y1posts}
\end{figure}

The resulting fitted correlations and posterior distributions are shown respectively in Figures \ref{fig:Y1corr} and \ref{fig:Y1posts}, while the corresponding BAO parameters are summarized in Table \ref{tab:Y1results}. The unblinded results are fully consistent with the DESI DR1 Lyman-$\alpha$ BAO baseline reported in \cite{Y1-BAO}, validating our implementation of the analysis pipeline. The unblinded contour is obtained by applying the same analysis pipeline to the original (unblinded) catalog and is shown here as a consistency check with the published DESI DR1 results. The blinded results remain consistent with those obtained for the synthetic catalogs, supporting the robustness of the blinding scheme under realistic observational conditions.

\begin{table}[tbp]
\centering
\begin{tabular}{| l | c | c | c | c | c |}
    \hline
\multirow{2}{*}{Catalog}  & \multirow{2}{*}{$\frac{\chi^2}{N_{data}-N_{pars}}$} & \multicolumn{2}{c|}{$\alpha_{\parallel}$} & \multicolumn{2}{c|}{$\alpha_{\perp}$} \\
    \cline{3-6}
    &  & Meas. & Error & Meas. & Error\\ \hline
Baseline	   & 1.011 & 0.9883 & 0.0196 & 1.0129 & 0.0233\\ \hline
Y1 unblind & 1.009 & 0.9882 & 0.0197 & 1.0133 & 0.0231\\ \hline
Y1 blind	   & 1.016 & 0.9491 & 0.0176 & 0.9782 & 0.0234\\ \hline
\end{tabular}
\caption{Best-fit values and uncertainties for the $\alpha_{\parallel}$ and $\alpha_{\perp}$ parameters, as well as the reduced chi-squared values, obtained with Vega on our Y1 catalog. The baseline values, taken from \cite{Y1-BAO}, are included for comparison and a consistency check.}
\label{tab:Y1results}
\end{table}

\subsection{Summary of robustness, limitations, and computational performance}
\label{sec:robustness}

\begin{figure}[t]
\centering 
\includegraphics[width=.8\textwidth]{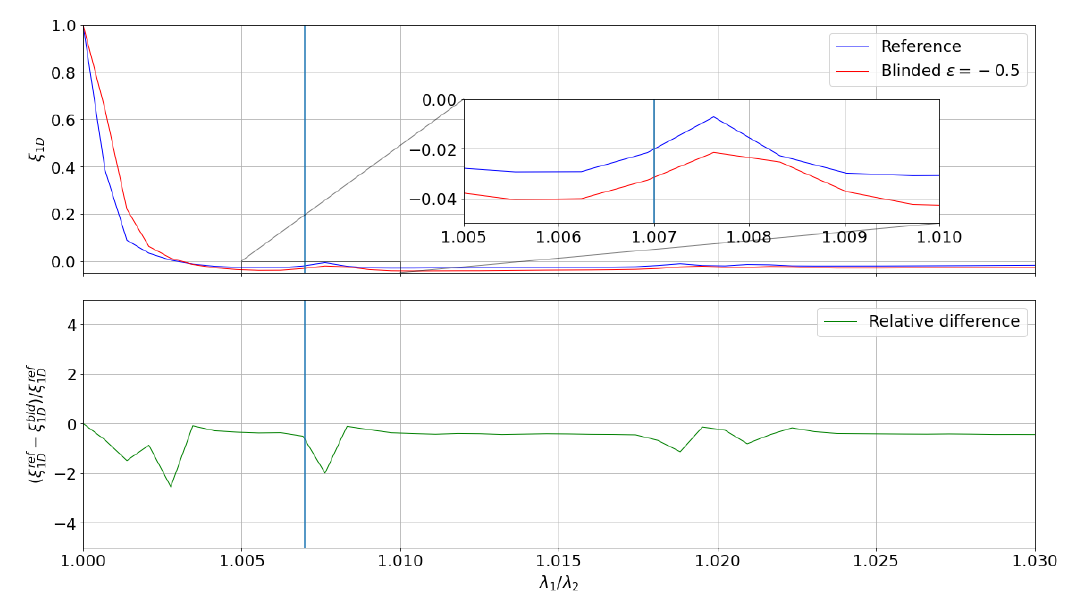}
\caption{One-dimensional Lyman-$\alpha$ correlation function. The relative difference (green) indicates that the metal-line peaks are shifted as a consequence of the blinding method. A shift can be seen for the peak at $\lambda_1/\lambda_2\sim 1.007$, which can give a hint on the direction of the applied blinding.}
\label{fig:1Dcorr}
\end{figure}

The presence of metal absorption lines introduces a potential avenue for partial unblinding through the one-dimensional correlation function. This arises from metal-induced features whose rest-frame wavelength positions are known from atomic physics and therefore do not follow the cosmological remapping, as illustrated in Figure \ref{fig:1Dcorr}.

The apparent displacement of the metal-induced features does not correspond to a physical displacement between the metal absorbers and the hydrogen gas. Rather, it results from assigning coordinates to all absorption features using the Ly-$\alpha$ rest-frame wavelength when applying the cosmological remapping. In principle, known metal features, such as the Si\,III doublet, could therefore provide information about the blinding transformation through their shifted relative positions. However, reconstructing the full transformation from these features is not straightforward, since it would require inverting the redshift transformation that produced the blinded coordinates, using the displaced metal correlation peaks as a reference to the Ly-$\alpha$ peak. Thus, the metal features can provide information about the direction of the blinding, but do not provide a direct determination of the blinding parameters. This potential source of information can be further reduced by applying additional blinding steps at the posterior or correlation-function level.

Another validation step is to examine the Distortion Matrix (DM), which accounts for distortions in the measured correlation functions introduced primarily by the continuum fitting process. Since the same continuum-fitting procedure is applied to the blinded and unblinded data, the dominant source of the distortion is unchanged by the blinding. Therefore, we do not expect the blinding procedure itself to introduce significant changes in the DM. However, the distortion matrix is calculated explicitly from the delta files and depends on the pixel weights used in this calculation. Since the rebinning associated with the blinding induces small changes in these weights, it can lead to small corresponding changes in the calculated DM. These changes in the weights also affect the numerical calculation of the DM, including its convergence and computation time.

\begin{figure}[t]
\centering 
\includegraphics[width=.8\textwidth]{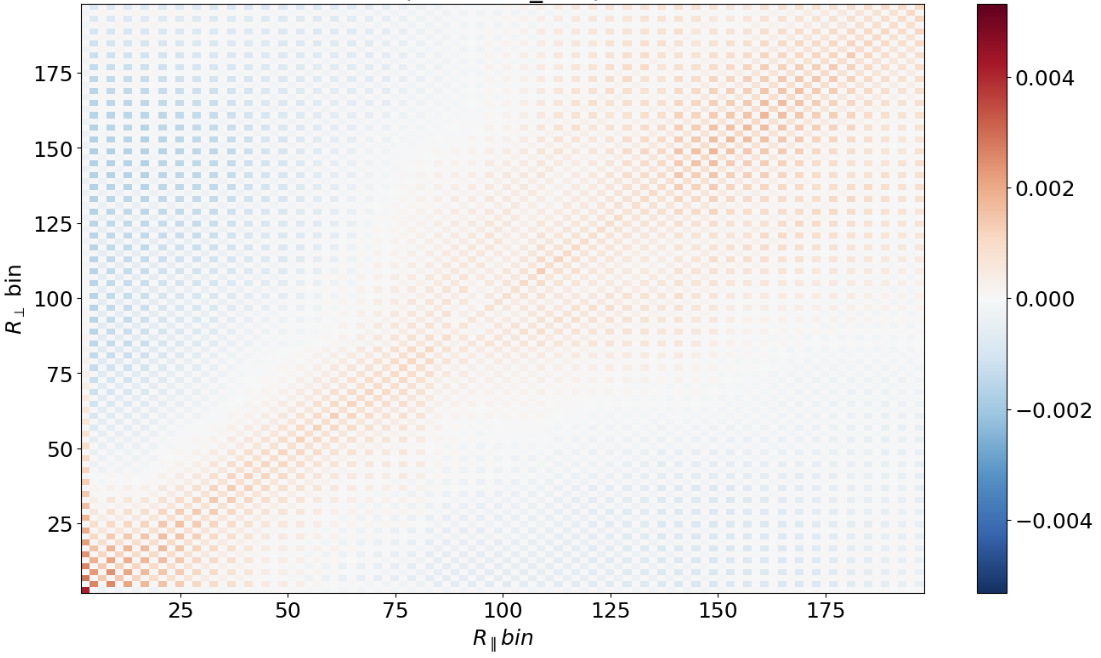}
\caption{Relative difference between the unblinded and blinded Distortion matrices for the Y1 dataset, calculated as $(DM - DM')/DM$. The observed deviations are  below 0.6\% of the amplitude of the unblinded matrix, indicating that the blinding introduces only very small modifications to the distortion matrix.}
\label{fig:DM}
\end{figure}

These effects are known to be accurately modeled, with their corresponding impact on the BAO peak position discussed in \cite{Busca2025}. The DM is therefore a useful diagnostic of the small effects introduced by the blinding procedure. A larger impact of the blinding on the weights would lead to more significant modifications of the matrix, as well as changes in the convergence of its calculation. In contrast, with our fiducial blinding scheme the DM shows only minimal modifications, as seen in Figure \ref{fig:DM}. The deviations are slightly stronger along the diagonal and weaker away from it, reflecting the mild changes in the weights induced by the rebinning procedure. For $\epsilon=-0.05$, the maximum relative deviation remains below $0.6\%$. For our fiducial blinding scheme, these changes do not lead to significant convergence issues. These results are consistent with the negligible differences observed in the correlation functions and the inferred BAO parameters.

The impact of the blinding on the covariance matrix depends on how the covariance is estimated. When the covariance is obtained directly from the data, for example using a jackknife or bootstrap procedure, the blinding is applied before the covariance estimation and its effects are therefore naturally propagated into the resulting covariance matrix. In contrast, when the covariance is derived from an independent set of mocks, the cosmology assumed for those mocks is not modified by the blinding. In this case, the fiducial cosmology used to generate the mocks should not be too far from either the blinded or unblinded cosmology, so that the covariance remains representative of the data under both descriptions. This provides an additional practical motivation for restricting the blinding parameters to moderate values.

We have tested the blinding procedure for values of $\epsilon$ ranging from $-0.05$ to $0.1$. Over this range, we find no significant differences in the correlation functions or inferred BAO parameters, while the modifications to the distortion matrix remain below the percent level. These tests indicate that the procedure is robust over the range explored here. We do not interpret this range as a universal upper limit on the blinding shift. Larger values of $\epsilon$ can increase the effects associated with rebinning and the resulting changes in pixel weights, and can also increase the mismatch between the cosmology of the blinded data and that used to construct an external mock-based covariance. The blinding parameters should therefore be chosen to be sufficiently large to hide the cosmological result, while remaining moderate enough to avoid unnecessary changes to the statistical properties of the data.

From a theoretical perspective, the blinding transformation preserves the information content of the delta field. The main source of potential changes in the posterior shape is therefore the rebinning of the forest, which slightly modifies the distribution of pixels and weights. However, quantities such as weights and distortion matrices show only percent-level variations, without significant changes in the inferred cosmological parameters.

Finally, the blinding prescription is computationally efficient, scaling linearly with the number of quasars. In practice, the additional computational cost corresponds to an increase of approximately 5\% in wall-clock time relative to the unblinded pipeline. This increase mainly arises from computing the mapping function and applying it to all pixels in the forest. Since the additional cost is small compared to the rest of the pipeline, no dedicated parallelization of the blinding was implemented and blinded and unblinded runs use equivalent computational resources. Since the method is modular, it can also be implemented independently of PICCA as a standalone preprocessing step.

\section{Conclusions}
\label{sec:conclusions}

We developed a catalog-level blinding methodology for the Lyman-$\alpha$ analysis based on a modification to the Alcock--Paczynski test, which is the standard methodology for galaxy clustering within DESI \cite{brieden} and has been validated for DESI analyses in \cite{validDR1, validDR2}. The prescription is based on a shift in the observer-frame wavelength of the forest delta field using the mapping between two different expansion histories. We verified the methodology using catalogs with not only a large number of low-noise spectra but also with a more realistic DESI Year 1 (Y1) sample, and we applied the methodology to the full Y1 dataset from DESI. There are mild changes in the blinded field as a consequence of the wavelength rebinning needed for the delta field to have a constant spacing in either a linear or a logarithmic scale. These changes slightly modify the pixel weights, which are propagated into the distortion matrix and the subsequent correlation functions. For our fiducial blinding scheme, the maximum relative modification of the distortion matrix remains below $0.6\%$. The BAO peak positions fitted along and perpendicularly to the line of sight in the blinded catalogs match the expected values obtained from the shifted cosmology.

In the case of the low-noise mocks, the shape of the posteriors remains almost intact, with a shifted central value that matches the theoretical expectation within one sigma, for a reasonable change of $5\%$ in the expansion history. For the Y1 catalogs, the change in the posterior's width is not negligible but also smaller than $5\%$, indicating a mild dependency on the blinding, likely associated with the rebinning step. We find no significant changes in the correlation functions or inferred BAO parameters over the range of blinding parameters explored in this work.

The proposed blinding scheme is applied after extracting the forest fluctuations, and it only takes a fraction of the time of this field extraction. We believe that this is the earliest step in the Lyman-$\alpha$ analysis pipeline where one can apply blinding, since the continuum fitting process allows one to separate the QSO contribution from the absorption features, both of which are shifted separately in the blinding process due to their distinct redshift origin. Furthermore, applying the scheme at this stage ensures that there is no extra bias on the systematics, since it happens after masking of DLAs, BALs, etc. The rebinning required by the blinding produces mild changes in the pixel weights, which are propagated into the distortion matrix and the subsequent correlation functions. We find that performing a further weight iteration after the blinding produces dramatic and unrealistic changes in the weights, and therefore this procedure is not used.

We show that the methodology presented here can indeed be considered as a sensible blinding option. When applied to real data from the first year of DESI, we show that the effect of metal contamination and Lyman-$\beta$ on the 3D auto- and cross-correlation functions is negligible and does not affect the algorithm's performance to hide the underlying cosmology, although metal absorption features in the 1D correlation function can provide information about the direction of the cosmological shift. This can be addressed by applying a secondary blinding prescription to the 1D correlation.

The covariance is naturally affected through the blinding when it is estimated directly from the data, while mock-based covariances require the mock cosmology to remain sufficiently close to the blinded and unblinded cosmologies. We therefore tested the procedure over $-0.05\leq\epsilon\leq0.1$, finding no significant impact on the correlation functions or inferred BAO parameters.

Finally, it is important to emphasize that this blinding technique enables a consistent analysis of cross-correlations between the Lyman-$\alpha$ forest and other tracers, provided that they share the same AP blinding. This is demonstrated in our Ly$\alpha\times$QSO cross-correlation results and opens the door to combining data from different tracers in the same or different experiments. Exploring a full-shape analysis in future work could also offer additional insights, provided that the associated challenges with redshift-space distortions are addressed with sufficient rigor. Furthermore, given the satisfactory results obtained for the Y1 catalog and the low-noise mocks, we expect the method to be applicable to later data releases. Its computational cost scales linearly with the number of quasars, making it practical for larger datasets, and the improved statistics should yield smaller errors, bringing the analysis closer to our idealized case of noiseless mocks.

\section{Data Availability}

Data from the plots in this paper are available on Zenodo as part of DESI's Data Management
Plan (\url{https://doi.org/10.5281/zenodo.21141840}). The data used in this analysis is public along the Data Release 1 (details in \url{https://data.desi.lbl.gov/doc/releases/dr1/}).

\section{Acknowledgments}
GPS, SB and GN acknowledge the support of DAIP-UG and SECIHTI, especially through the graduate study fellowships, grant "Estancias Posdoctorales por Mexico 2023(1)" No. BP-PA-20230822185725045-5723543, and grants "Ciencia Básica y de Frontera" No. 102958 and CBF-2025-I-2795. The authors also thank the Instituto Avanzado de Cosmología and the DCI-UG DataLab for actions and resources that directly impacted this work. AFR acknowledges financial support from the Spanish Ministry of Science and Innovation through the PID2024-159420NB-C41 project and the ``Excelencia Severo Ochoa'' program (CEX2024-001441-S from MICIU AEI 10.13039/501100011033) and the European Union through the ERC Consolidator Grant program (COSMO-LYA, grant agreement 101044612). IFAE is partially funded by the CERCA program of the Generalitat de Catalunya.

This material is based upon work supported by the U.S. Department of Energy (DOE), Office of Science, Office of High-Energy Physics, under Contract No. DE–AC02–05CH11231, and by the National Energy Research Scientific Computing Center, a DOE Office of Science User Facility under the same contract. Additional support for DESI was provided by the U.S. National Science Foundation (NSF), Division of Astronomical Sciences under Contract No. AST-0950945 to the NSF’s National Optical-Infrared Astronomy Research Laboratory; the Science and Technology Facilities Council of the United Kingdom; the Gordon and Betty Moore Foundation; the Heising-Simons Foundation; the French Alternative Energies and Atomic Energy Commission (CEA); the Secretariat of Science, Humanities, Technology and Innovation (SECIHTI) of Mexico; the Ministry of Science, Innovation and Universities of Spain (MICIU/AEI/10.13039/501100011033), and by the DESI Member Institutions: \url{https://www.desi.lbl.gov/collaborating-institutions}. Any opinions, findings, and conclusions or recommendations expressed in this material are those of the author(s) and do not necessarily reflect the views of the U. S. National Science Foundation, the U. S. Department of Energy, or any of the listed funding agencies.

The authors are honored to be permitted to conduct scientific research on I'oligam Du'ag (Kitt Peak), a mountain with particular significance to the Tohono O’odham Nation.

\appendix
\section{PICCA implementation} 
\label{sec:PICCA}

In practice, we implement the AP blinding prescription in the set of tools for the Lyman-$\alpha$ algorithm called PICCA\footnote{\href{ttps://github.com/igmhub/picca/}{ttps://github.com/igmhub/picca/}}. This set of connected codes has been widely used for Lyman-$\alpha$ forest analyses in the eBOSS and, more recently, the DESI collaborations. In detail, the PICCA suite covers several end-to-end analysis tools for IGM analysis, such as the extraction of overdensity fields, the 1D power spectrum, the 1D and 3D two-point auto and cross correlation functions, covariance matrices and model fittings for correlation functions.

\begin{figure}[t]
\centering 
\includegraphics[width=.8\textwidth]{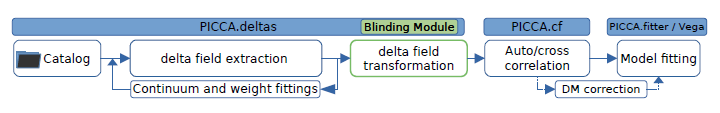}
\caption{PICCA workflow for computing correlation functions. The blinding is applied after the iterative process of the delta field extraction and before the calculation of correlation functions.}
\label{fig:PICCAworkflow}
\end{figure}

The pipeline of PICCA for the calculation of the AP test parameters (see eq. \eqref{eq:AlphaDef}) using the correlation functions is depicted in Figure \ref{fig:PICCAworkflow}. One often starts with a script, named picca$\_$delta$\_$extraction, which extracts the forest delta field for Lyman-$\alpha$ (defined by default in the wavelength range 1060\AA-1200\AA) or Lyman-$\beta$ (with no default value, but which we assume 920\AA-1020\AA). To use picca$\_$delta$\_$extraction, an input catalog is needed, composed of positions, redshifts and spectra of a quasar sample above $z = 2.1$. The spectra are read from an additional set of files, which are organized in HEALPIX equal area sky pixel\footnote{See \href{https://healpix.jpl.nasa.gov}{https://healpix.jpl.nasa.gov} for further details.}. Moreover, the user can specify several options in the code, including: the wavelength range of the forests, the threshold of quality for the QSO redshift estimation, sky lines, dust maps and galactic extinction for cleansing the spectra, reject forest with BALs (Broad absorption lines), mask DLAs (Damped Lyman-$\alpha$ systems) and select from one to multiple DESI expositions for re-observed QSOs.

While working with the forest's data, which would have a resolution given by the instrument, PICCA resamples the data points to a homogeneously spaced grid (linear or logarithmic in scale, with $\Delta\log(\lambda)\sim 10^{-4}$ before extracting the overdensity field. The code fits the continuum using a particular model, which we describe shortly, and then saves the delta field as separate files in pixels of equal sky area. These delta files are then used as inputs in several PICCA scripts that can calculate the 1D power spectrum, 1D correlation function, among others. To our interest, the PICCA$\_$cf script calculates the 3D correlation functions. Following the 3D correlation function, one can fit using a $\chi^2$ minimizer or do a Bayesian analysis to infer the values for several quantities of interest. This can be done using PICCA or using the most recent package for the correlation function modeling Vega.

Given that our blinding code depends on the script picca$\_$delta$\_$extraction and the different assumptions made within, we would like to describe in more details how the output delta field is obtained. To calculate the overdensity field, first, we need to estimate the spectrum of each quasar prior to the absorptions on it from the intergalactic media and other objects along the forest's trajectory. This "clean" spectrum is usually referred to as the Continuum and, once estimated, it can be subtracted from the measured flux to obtain the delta field. At the level of PICCA, the standard pipeline includes an algorithm that allows the synchronous calculation of both the delta field and the continuum fitting, in a process called delta extraction. Let us describe how this works in a more formal manner: the quantity $\Delta(\lambda)$, from which the overdensity field is later constructed, is computed for the i-th QSO as
\begin{equation}
 \Delta^i = \frac{f^i}{\bar{F}C^i}.
\end{equation}
This equation does not correspond to the desired delta field, $\delta^i$, of equation (\ref{eq:delta}) as it has not yet been centered at zero; this will be done in a later step. The function $f^i$ represents the flux of the QSO, while the denominator represents the continuum of the QSO and is separated in two parts: $\bar{F}$ is a base mean function shared by all QSO's emission spectra and $C^i$ is a polynomial correction to the base form for the i-th QSO. The latter is calculated in PICCA pipeline as a first order polynomial and thus represents a spectral slope. The product $\bar{F}C^i$ is estimated as a whole by an iterative algorithm based on variance minimization, since the true continuum $C$ is challenging to obtain on its own. In order to do this, PICCA calculates and corrects $\bar{C}$ on each iteration as
\begin{equation}
 \bar{F}(\lambda_{\mathrm{rest}}) = \sum_i \Delta^i(\lambda_{\mathrm{rest}})\omega^i(\lambda_{\mathrm{rest}}),
\end{equation}
where the superindex $i$ runs over the QSOs of the sample. However, notice that in the first iteration there are no deltas to calculate $\bar{F}$, and so it is assumed as equal to unity and the weights for each QSO are the inverse variance weighting as provided by the instrument. For the rest of the iterations $\omega$ becomes the inverse of the calculated variance. As $\bar{F}$ is the common part of the emission spectra for all the QSOs of the catalog, it is calculated in the restframe. Individual correction factors for the emission spectra, $C^i$, are represented in the observed frame by
\begin{equation}
 C^i(\lambda_{\mathrm{obs}}) = a^i+b^i\log(\lambda_{\mathrm{obs}})
 \label{eq:C_i}
\end{equation}
The parameters $a^i,\ b^i$ are to be obtained by minimizing the expression
\begin{equation}
 \chi^2 = \omega(\lambda_{\mathrm{obs}})\left[f^i(\lambda_{\mathrm{obs}})-\bar{F}(\lambda_{\mathrm{obs}})C^i(\lambda_{\mathrm{obs}})\right]^2-1.
\end{equation}

The pipeline variance used in the first iteration and calculated in terms of the inverse variance weighting of the instrument is
\begin{equation*}
 \left(\sigma_{\mathrm{pip}}^2\right)^i = \frac{1}{\omega_{\mathrm{ivar}}^i\left(C^i\bar{F}\right)^2}
\end{equation*}
In the following iterations the variance is calculated minimizing the more complex expression:
\begin{equation}
 (\sigma^2)^j = \eta^j\left\langle\sigma_{\mathrm{pip}}^2\right\rangle^j+(\sigma_{\mathrm{LSS}}^2)^j+\frac{\epsilon^j}{\left\langle\sigma_{\mathrm{pip}}^2\right\rangle^j},
\label{eq:varianceiterations}
\end{equation}
where $j$ runs over a series of equally spaced wavelength bins that covers the whole observed frame region of the spectrometer for the lyman-$\alpha$ science (usually 20 bins from 3600 to
5500 \AA). The term $\left\langle\sigma_{\mathrm{pip}}^2\right\rangle^j$ is the average variance of all QSOs with flux within the j-th wavelength bin. The variance $(\sigma^2)^j$ is minimized independently for each j-th bin by varying three variables: the correction to the instrument weighting $\eta$ that is proportional to the pipeline variance.
The $\sigma_{\mathrm{LSS}}$ factor adds variances that are independent of $(\sigma^2)^j$, such as the effect of Large Scale Structure (hence the label "LSS") or variance added by rebinning the spectra. Finally, the fudge factor $\epsilon$ adds variance to high signal to noise QSOs due to the lack of diversity in the continuum model for the emission spectra. Once the three variance parameters are calculated for each wavelength bin, their new weights are used to estimate the $C^i, \bar{F}$ for the next iteration. At the same time, the mean for the $\Delta^i$ expression is removed using an observed frame stack $\Delta-\left\langle\Delta\right\rangle$, where the average is given by
\begin{equation}
 \left\langle\Delta\right\rangle = \frac{\sum_i\Delta^i(\lambda_{\mathrm{obs}})\omega^i(\lambda_{\mathrm{obs}})}{\sum_i\omega^i(\lambda_{\mathrm{obs}})}.
\end{equation}
After convergence, the final overdensity function is calculated within the observed frame as
\begin{equation}
 \delta^i = \frac{f^i}{\bar{F}C^i}-\left\langle\Delta\right\rangle.
\end{equation}
Once the overdensity field is calculated, one may want to do a statistical analysis. Prior to this step, however, there is a set of corrections to the field that can be implemented through PICCA. The first one is a correction to the zero-centering of each forest, that can be biased because the true continuum of the QSO is not known. The distortion matrix script statistically processes a smaller forest sample within the catalog to statistically calculate and reduce this bias. Another tool is the metallic distortion matrix process, which calculates the effect that each of the metallic absorptions would have over the correlation function. These processes are calculated directly from the deltas and are used in the last steps of the correlation function estimation.

For this paper we calculate the two-point correlation function (2PCF) and 1D correlation function, with each of them having its own script included within the PICCA package. In order to calculate the 2PCF, the overdensity field is mapped from wavelengths to the redshifts at which each absorption is produced, and then, to a comoving distance space associated to the redshifts using a fiducial cosmology (i.e. Planck 2015 \cite{54}). Note that the blinding can be directly implemented in this mapping process, however, it is the most obvious part and it would be easy to unblind. Furthermore, it leaves the delta field unblinded if analyzed with another set of tools or by another PICCA script.

Once the field is mapped to the comoving distance space, the correlation function is calculated for each HEALPIXs and for each forest set between a maximum distance (usually $200\ Mpc/h$). Further technical details of the calculation of the 2PCF and 1D correlation are beyond the scope of this paper (see \cite{DESI2024.II.KP3} for more details). Once the correlation function is calculated, the previously calculated distortion matrix can be used to lower the bias induced to the correlation function due to bad normalization of the QSO spectra during the delta extraction.

To follow up the calculation of the AP parameters, once the 2PCF is estimated, one can apply a fitting algorithm to it and use the result to infer cosmological information. PICCA fits the Alcock-Paczynski anisotropic clustering parameters in Eq. \eqref{eq:AlphaDef}, as well as bin sizes, small scale non-linear effects, growth rate of cosmic structure, peak smoothening and parameters to reduce bias from different sources such as high column density (HCD), BALS and metal absorptions through a metal distortion matrix, among others. For the blinded case there are input parameters that have to be modified. The minimum and maximum comoving distance parameters for the correlation
function calculation have to be modified to higher values ($5\ Mpc/h$ for the case $\Omega_m = 0.95$), in order to include the contribution of high redshift QSO that are moved beyond the border (or to
include low redshift QSOs shifted below the lower limit for $\Omega'_m > \Omega_m$).

To summarize, the blinding process affects geometrically the comoving distances of each point of the overdensity field according to the change of the dark matter content of the blinded cosmology. Hence, the mean redshift $z_{\mathrm{eff}}$  of the catalog must be modified accordingly. The new value can be calculated taking the mean of the shifted redshifts ($z'_{\mathrm{QSO}}$), or approximated using eq. \eqref{eq:f(z)} with the unblinded value as input. The approximation holds exact if the sample is homogeneous across the redshift range of the sample and it is used while blinding galaxy clustering \cite{brieden}; however, this approximation does not hold for the distribution of the datasets hereby mentioned. We calculate the mean redshift using the former and more rigorous method to avoid introducing errors from the approximation.


\section{Author Affiliations}
\label{sec:affiliations}

\noindent \hangindent=.5cm $^{1}${Departamento de F\'{\i}sica, DCI-Campus Le\'{o}n, Universidad de Guanajuato, Loma del Bosque 103, Le\'{o}n, Guanajuato C.~P.~37150, M\'{e}xico}

\noindent \hangindent=.5cm $^{2}${Instituto Avanzado de Cosmolog\'{\i}a A.~C., San Marcos 11 - Atenas 202. Magdalena Contreras. Ciudad de M\'{e}xico C.~P.~10720, M\'{e}xico}

\noindent \hangindent=.5cm $^{3}${Institute for Theoretical Particle Physics and Cosmology (TTK), RWTH Aachen University, Sommerfeldstr. 16, D-52056 Aachen, Germany}

\noindent \hangindent=.5cm $^{4}${Instituci\'{o} Catalana de Recerca i Estudis Avan\c{c}ats, Passeig de Llu\'{\i}s Companys, 23, 08010 Barcelona, Spain}

\noindent \hangindent=.5cm $^{5}${Institut de Ci\`encies del Cosmos (ICCUB), Universitat de Barcelona (UB), c. Mart\'i i Franqu\`es, 1, 08028 Barcelona, Spain.}

\noindent \hangindent=.5cm $^{6}${Institut de F\'{i}sica d’Altes Energies (IFAE), The Barcelona Institute of Science and Technology, Edifici Cn, Campus UAB, 08193, Bellaterra (Barcelona), Spain}

\noindent \hangindent=.5cm $^{7}${Lawrence Berkeley National Laboratory, 1 Cyclotron Road, Berkeley, CA 94720, USA}

\noindent \hangindent=.5cm $^{8}${Department of Physics, Boston University, 590 Commonwealth Avenue, Boston, MA 02215 USA}

\noindent \hangindent=.5cm $^{9}${Dipartimento di Fisica ``Aldo Pontremoli'', Universit\`a degli Studi di Milano, Via Celoria 16, I-20133 Milano, Italy}

\noindent \hangindent=.5cm $^{10}${INAF-Osservatorio Astronomico di Brera, Via Brera 28, 20122 Milano, Italy}

\noindent \hangindent=.5cm $^{11}${Department of Physics \& Astronomy, University College London, Gower Street, London, WC1E 6BT, UK}

\noindent \hangindent=.5cm $^{12}${Instituto de F\'{\i}sica, Universidad Nacional Aut\'{o}noma de M\'{e}xico,  Circuito de la Investigaci\'{o}n Cient\'{\i}fica, Ciudad Universitaria, Cd. de M\'{e}xico  C.~P.~04510,  M\'{e}xico}

\noindent \hangindent=.5cm $^{13}${Department of Astronomy \& Astrophysics, University of Toronto, Toronto, ON M5S 3H4, Canada}

\noindent \hangindent=.5cm $^{14}${Department of Physics \& Astronomy and Pittsburgh Particle Physics, Astrophysics, and Cosmology Center (PITT PACC), University of Pittsburgh, 3941 O'Hara Street, Pittsburgh, PA 15260, USA}

\noindent \hangindent=.5cm $^{15}${University of California, Berkeley, 110 Sproul Hall \#5800 Berkeley, CA 94720, USA}

\noindent \hangindent=.5cm $^{16}${Departamento de F\'isica, Universidad de los Andes, Cra. 1 No. 18A-10, Edificio Ip, CP 111711, Bogot\'a, Colombia}

\noindent \hangindent=.5cm $^{17}${Observatorio Astron\'omico, Universidad de los Andes, Cra. 1 No. 18A-10, Edificio H, CP 111711 Bogot\'a, Colombia}

\noindent \hangindent=.5cm $^{18}${Institut d'Estudis Espacials de Catalunya (IEEC), c/ Esteve Terradas 1, Edifici RDIT, Campus PMT-UPC, 08860 Castelldefels, Spain}

\noindent \hangindent=.5cm $^{19}${Institute of Cosmology and Gravitation, University of Portsmouth, Dennis Sciama Building, Portsmouth, PO1 3FX, UK}

\noindent \hangindent=.5cm $^{20}${Institute of Space Sciences, ICE-CSIC, Campus UAB, Carrer de Can Magrans s/n, 08913 Bellaterra, Barcelona, Spain}

\noindent \hangindent=.5cm $^{21}${University of Virginia, Department of Astronomy, Charlottesville, VA 22904, USA}

\noindent \hangindent=.5cm $^{22}${Fermi National Accelerator Laboratory, PO Box 500, Batavia, IL 60510, USA}

\noindent \hangindent=.5cm $^{23}${Institut d'Astrophysique de Paris. 98 bis boulevard Arago. 75014 Paris, France}

\noindent \hangindent=.5cm $^{24}${IRFU, CEA, Universit\'{e} Paris-Saclay, F-91191 Gif-sur-Yvette, France}

\noindent \hangindent=.5cm $^{25}${Center for Cosmology and AstroParticle Physics, The Ohio State University, 191 West Woodruff Avenue, Columbus, OH 43210, USA}

\noindent \hangindent=.5cm $^{26}${Department of Physics, The Ohio State University, 191 West Woodruff Avenue, Columbus, OH 43210, USA}

\noindent \hangindent=.5cm $^{27}${The Ohio State University, Columbus, 43210 OH, USA}

\noindent \hangindent=.5cm $^{28}${Department of Physics, University of Michigan, 450 Church Street, Ann Arbor, MI 48109, USA}

\noindent \hangindent=.5cm $^{29}${University of Michigan, 500 S. State Street, Ann Arbor, MI 48109, USA}

\noindent \hangindent=.5cm $^{30}${Department of Physics, The University of Texas at Dallas, 800 W. Campbell Rd., Richardson, TX 75080, USA}

\noindent \hangindent=.5cm $^{31}${NSF NOIRLab, 950 N. Cherry Ave., Tucson, AZ 85719, USA}

\noindent \hangindent=.5cm $^{32}${Sorbonne Universit\'{e}, CNRS/IN2P3, Laboratoire de Physique Nucl\'{e}aire et de Hautes Energies (LPNHE), FR-75005 Paris, France}

\noindent \hangindent=.5cm $^{33}${Departament de F\'{i}sica, Serra H\'{u}nter, Universitat Aut\`{o}noma de Barcelona, 08193 Bellaterra (Barcelona), Spain}

\noindent \hangindent=.5cm $^{34}${Department of Astronomy, The Ohio State University, 4055 McPherson Laboratory, 140 W 18th Avenue, Columbus, OH 43210, USA}

\noindent \hangindent=.5cm $^{35}${Department of Physics and Astronomy, University of Waterloo, 200 University Ave W, Waterloo, ON N2L 3G1, Canada}

\noindent \hangindent=.5cm $^{36}${Perimeter Institute for Theoretical Physics, 31 Caroline St. North, Waterloo, ON N2L 2Y5, Canada}

\noindent \hangindent=.5cm $^{37}${Waterloo Centre for Astrophysics, University of Waterloo, 200 University Ave W, Waterloo, ON N2L 3G1, Canada}

\noindent \hangindent=.5cm $^{38}${Instituto de Astrof\'{i}sica de Andaluc\'{i}a (CSIC), Glorieta de la Astronom\'{i}a, s/n, E-18008 Granada, Spain}

\noindent \hangindent=.5cm $^{39}${Departament de F\'isica, EEBE, Universitat Polit\`ecnica de Catalunya, c/Eduard Maristany 10, 08930 Barcelona, Spain}

\noindent \hangindent=.5cm $^{40}${Department of Physics and Astronomy, Sejong University, 209 Neungdong-ro, Gwangjin-gu, Seoul 05006, Republic of Korea}

\noindent \hangindent=.5cm $^{41}${CIEMAT, Avenida Complutense 40, E-28040 Madrid, Spain}

\noindent \hangindent=.5cm $^{42}${Space Telescope Science Institute, 3700 San Martin Drive, Baltimore, MD 21218, USA}

\bibliographystyle{JHEP}  
 \bibliography{references.bib}

\end{document}